\documentclass[10pt]{article}
\usepackage{bm,latexsym,amsmath,amsthm,amssymb}
\usepackage{graphicx,indentfirst}
\usepackage[numbers,sort&compress]{natbib}
\date{}
\begin{document}
\title{\bf On the Cauchy problem for a higher-order $\mu$-Camassa-Holm equation}
\author{Feng Wang\thanks{Corresponding author.\newline
\mbox{}\qquad E-mail: wangfeng@xidian.edu.cn (F. Wang); fqli@dlut.edu.cn (F. Li); zhijun.qiao@utrgv.edu (Z. Qiao); }\\
\small School of Mathematics and Statistics, Xidian University,
    Xi'an 710071, PR China\\[3pt]
Fengquan Li\\
\small School of Mathematical Sciences, Dalian University of Technology,
Dalian 116024, PR China\\[3pt]
Zhijun Qiao\\
\small School of Mathematical \& Statistical Sciences,\\
\small University of Texas-Rio Grande Valley, Texas 78539, USA}
\date{}
\maketitle \baselineskip 3pt
\begin{center}
\begin{minipage}{130mm}
{{\bf Abstract.} In this paper, we study the Cauchy problem of a higher-order $\mu$-Camassa-Holm equation. We first establish the Green's function of $(\mu-\partial_{x}^{2}+\partial_{x}^{4})^{-1}$ and local well-posedness for the equation in Sobolev spaces $H^{s}(\mathbb{S})$, $s>\frac{7}{2}$. Then we provide the global existence results for strong solutions and weak solutions. Moreover, we show that the solution map is non-uniformly continuous in $H^{s}(\mathbb{S})$, $s\geq 4$.
Finally, we prove that the equation admits single peakon solutions.

}

\end{minipage}
\end{center}

\baselineskip=15pt

\section{Introduction}
\label{intro}

The Camassa-Holm equation
\begin{align*}\label{1.1}
m_{t}+2mu_{x}+m_{x}u=0, \quad m=(1-\partial_{x}^{2})u
 \tag{1.1}
\end{align*}
was introduced in \cite{ch93} to model the unidirectional propagation of shallow water waves over a flat bottom. $u(t, x)$ represents the fluid velocity at time $t$ and in the spatial direction $x$. It is a re-expression of the geodesic flow both on the diffeomorphism group of the circle \cite{ck02} and on the Bott-Virasoro group \cite{k07}. Eq.(\ref{1.1}) has a bi-Hamiltonian structure \cite{ff81} and is completely integrable \cite{ch93, c98}. Moreover, it has been extended to an entire integrable hierarchy including both negative and positive flows and shown to admit algebro-geometric solutions on a symplectic submanifold \cite{Qiao2003}.
The Cauchy problem of (\ref{1.1}), in particular its well-posedness, blow-up behavior and global existence, have been well-studied both on the real line and on the circle, e.g., \cite{brc07,c97,ce98,cm99,d01,cs00,ca00,cm00,lo00,xz00,ces98,ce00,himk09,himkm10}.
Eq.(\ref{1.1}) with weakly dissipative term was studied in \cite{wy09}.

Equation (\ref{1.1}) has been recently generalized into some $\mu$-versions and higher order forms. Khesin et al. in \cite{klm08} introduced a $\mu$-version of Camassa-Holm equation as follows
\begin{align*}\label{1.2}
m_{t}+2mu_{x}+m_{x}u=0, \quad m=(\mu-\partial_{x}^{2})u,
 \tag{1.2}
\end{align*}
where $u(t,x)$ is a time-dependent function on the unit circle $\mathbb{S}=\mathbb{R}/\mathbb{Z}$ and $\mu(u)=\int_{\mathbb{S}}udx$ denotes its mean.
This equation describes the propagation
of weakly nonlinear orientation waves in a massive nematic liquid crystal with external
magnetic filed and self-interaction. Moreover, Eq.(\ref{1.2}) is also an Euler equation
on $\mathcal{D}^{s}(\mathbb{S})$ and it describes the geodesic flow on $\mathcal{D}^{s}(\mathbb{S})$ with the right-invariant metric given at the identity by the inner product \cite{klm08}
$$
\begin{array}{l}
\langle f,g\rangle_{\mu}=\mu(f)\mu(g)+\int_{\mathbb{S}}f^{\prime}(x)g^{\prime}(x)dx~.
\end{array}
$$
In \cite{klm08,lmt10}, the authors showed that Eq.(\ref{1.2}) is bi-Hamiltonian and admits both cusped and smooth travelling wave solutions which are natural candidates for solitons.
The orbit stability of periodic peakons was studied in \cite{clel13}. A weakly dissipative $\mu$-Camassa-Holm equation was studied in \cite{ly14}.

For the higher order Camassa-Holm equation, \cite{ck03,chk09} considered the following equation
\begin{align*}\label{1.3}
m_{t}+2mu_{x}+m_{x}u=0, \quad m=\sum_{j=0}^{k}(-1)^{j}\partial_{x}^{2j}u,
 \tag{1.3}
\end{align*}
which describes exponential curves of the manifold of smooth orientation-preserving diffeomorphisms of the unit circle in the plane. In \cite{chk09}, Coclite et al. established the existence of global weak solutions and presented some invariant spaces under the action of the equation.
Tian et al. \cite{tzx11} investigated the global existence of strong solutions to Equation (\ref{1.3}) with $k=2$. Ding and Lv \cite{dingl10} studied the existence of global conservative solutions to (1.3). Recently, Coclite and Ruvo \cite{coclite17} showed  the convergence of the solution to (1.3). Ding et al. \cite{ding17,dingd17} discussed traveling solutions of (\ref{1.3}) and their evolution properties.

In this paper, we will consider a $\mu$-version of (\ref{1.3}) with $k=2$ as follows
\begin{align*}\label{1.4}
m_{t}+2mu_{x}+m_{x}u=0, \quad m=(\mu-\partial_{x}^{2}+\partial_{x}^{4})u,
 \tag{1.4}
\end{align*}
where $u(t,x)$ is a time-dependent spatially periodic function on the unit-circle $\mathbb{S}=\mathbb{R}/\mathbb{Z}$ and $\mu(u)=\int_{\mathbb{S}}udx$ denotes its mean.

We first give the Green's function of the operator
$(\mu-\partial_{x}^{2}+\partial_{x}^{4})^{-1}$ and local well-posedness of (\ref{1.4}).
Then we show the global existence of strong solutions to (\ref{1.4}).
Next, for any $T_{0}>0$ and $s\geq 4$, we prove that the data-to-solution map is H\"{o}lder continuous from any bounded subset of $H^{s}(\mathbb{S})$ into $C([0, T_{0}]; H^{r}(\mathbb{S}))$ with $0\leq r<s$, but is not uniformly continuous from any bounded subset of $H^{s}(\mathbb{S})$ into $C([0, T_{0}]; H^{s}(\mathbb{S}))$.
Motivated by the recent work \cite{cokar15}, we establish the existence of global weak solution in $H^{2}(\mathbb{S})$ without using an Ole\u{\i}nik-type estimate
(see \cite{xz00,chkar06}),
which is not easy to be verified in numerical experiment. Lastly, we show the existence of single peakon solutions.

We noticed that Mclachlan and Zhang \cite{mczh09} have studied another higher-order Camassa-Holm equation as follows
\begin{align*}\label{1.5}
m_{t}+2mu_{x}+m_{x}u=0, \quad m=(1-\partial_{x}^{2})^{k},
 \tag{1.5}
\end{align*}
which is derived as the Euler-Poincar\'{e} differential equation on the Bott-Virasoro group with respect to the $H^{k}$ metric. A $\mu$-version of (\ref{1.5}) with $k=2$, first proposed in \cite{eskol14}, was very recently studied in our recent paper \cite{wang17}, in which we also established the Green's function of the operator $(\mu-\partial_{x}^{2})^{-2}$ and showed it admits single peakon solutions, but they are completely different from the results in the present paper.

The rest of the paper is organized as follows. In Section 2, the Green's function of the operator
$(\mu-\partial_{x}^{2}+\partial_{x}^{4})^{-1}$ and local well-posedness for (\ref{1.4})
with initial data in $H^{s}(\mathbb{S}), s>\frac{7}{2}$, are established. In Section 3, we show the global existence of strong solutions. The H\"{o}lder continuity and non-uniform continuity of solution map for the equation is established in Section 4. In Section 5, we show the global existence of weak solutions. The existence of single peakon solutions is proved in Section 6.

\section{Preliminaries}

In this section, we will give the Green's function of the operator
$A_{\mu}^{-1}:=(\mu-\partial_{x}^{2}+\partial_{x}^{4})^{-1}$ and establish the local well-posedness for Eq. (1.4).

\subsection{Green's function}

To construct the peaked solutions in the last section, we need to investigate the Green's function of the operator $A_{\mu}^{-1}$. We denote the Fourier transform of $f$ by $\widehat{f}$.

For a periodic function $g$ on the circle $\mathbb{S}=\mathbb{R}/\mathbb{Z}$, we have
$$
\widehat{\mu(g)}(k)
=\int_{\mathbb{S}}\mu(g)(x)e^{-2\pi ikx}dx
=\mu(g)\int_{\mathbb{S}}e^{-2\pi ikx}dx
=\mu(g)\delta_{0}(k),
$$
where
$$
\begin{array}{l}
\delta_{0}(k)=\left\{\begin{array}{l}
1, \quad k=0, \\[3pt]
0, \quad k\neq 0.
\end{array}
\right.
\end{array}
$$
Since $\mu(g)=\widehat{g}(0)$, we have $\widehat{\mu(g)}(k)=\delta_{0}(k)\widehat{g}(k)$.
Thus,
$$
\widehat{A_{\mu}g}(k)
=\widehat{(\mu-\partial_{x}^{2}+\partial_{x}^{4})g}(k)
=[\delta_{0}(k)+(2\pi k)^{2}+(2\pi k)^{4}]\widehat{g}(k).
$$

If $g$ is the Green's function of the operator $A_{\mu}^{-1}$, that is, $g$ satisfies $A_{\mu}g=\delta(x)$,
then $[\delta_{0}(k)+(2\pi k)^{2}+(2\pi k)^{4}]\widehat{g}(k)=1$.
Thus,
$$
\begin{array}{rl}
g(x)
&=\sum_{k\in \mathbb{Z}}\widehat{g}(k)e^{2\pi ikx}
=\sum_{k\in \mathbb{Z}}\frac{1}{\delta_{0}(k)+(2\pi k)^{2}+(2\pi k)^{4}}e^{2\pi ikx}\\[5pt]
&=1+2\sum_{k=1}^{\infty}\frac{\cos 2\pi kx}{(2\pi k)^{2}+(2\pi k)^{4}}.
\end{array}
$$
By Weierstrass's criterion, we know the series
$$
\begin{array}{rl}
-\sum_{k=1}^{\infty}\frac{\sin 2\pi kx}{2\pi k+(2\pi k)^{3}}, \quad
-\sum_{k=1}^{\infty}\frac{\cos 2\pi kx}{1+(2\pi k)^{2}}
\end{array}
$$
uniformly converge in $[0, 1)\simeq \mathbb{S}$.
From Dirichlet's criterion, we know that the series
$$
\begin{array}{rl}
\sum_{k=1}^{\infty}\frac{(2\pi k)\sin 2\pi kx}{1+(2\pi k)^{2}}
\end{array}
$$
converge for any $x\in [0,1)\simeq\mathbb{S}$, and uniformly converge in any closed interval $[\alpha, \beta]\subset (0, 1)$. Thus, $g(x)$ is two-times continuously differentiable on $[0,1)\simeq \mathbb{S}$ and three-times continuously differentiable on $[\alpha, \beta]\subset (0, 1)$. It follows that $\|\partial_{x}^{i}g\|_{L^{\infty}(\mathbb{S})}<\infty$ $(i=0,1,2,3)$.

Note that
$$
\begin{array}{rl}
g(x)
&=1+2\sum_{k=1}^{\infty}\frac{\cos 2\pi kx}{(2\pi k)^{2}+(2\pi k)^{4}}\\[5pt]
&=1+2\sum_{k=1}^{\infty}\left(\frac{1}{(2\pi k)^{2}}-\frac{1}{1+(2\pi k)^{2}}\right)\cos 2\pi kx\\[5pt]
&=1+2\sum_{k=1}^{\infty}\frac{\cos 2\pi kx}{(2\pi k)^{2}}
-2\sum_{k=1}^{\infty}\frac{\cos 2\pi kx}{1+(2\pi k)^{2}}.
\end{array}
$$
Since the Green's functions of $(\mu-\partial_{x}^{2})^{-1}$ and $(1-\partial_{x}^{2})^{-1}$
are $g_{\mu}(x)=\frac{1}{2}(x-\frac{1}{2})^{2}+\frac{23}{24}$ and $g_{1}(x)=\frac{\cosh(x-\frac{1}{2})}{2\sinh(\frac{1}{2})}$ respectively, that is,
$$
g_{\mu}(x)=1+2\sum_{k=1}^{\infty}\frac{\cos 2\pi kx}{(2\pi k)^{2}},\quad
g_{1}(x)=1+2\sum_{k=1}^{\infty}\frac{\cos 2\pi kx}{1+(2\pi k)^{2}},
$$
the Green's function $g(x)$ is given by
$$
\begin{array}{rl}
g(x)
&=g_{\mu}(x)-g_{1}(x)+1\\[5pt]
&=\frac{1}{2}(x-\frac{1}{2})^{2}
-\frac{\cosh(x-\frac{1}{2})}{2\sinh(\frac{1}{2})}
+\frac{47}{24}, \quad x\in [0, 1)\simeq\mathbb{S},
\end{array}
$$
and is extended periodically to the real line, that is
$$
\begin{array}{rl}
g(x)
=\frac{1}{2}(x-[x]-\frac{1}{2})^{2}
-\frac{\cosh(x-[x]-\frac{1}{2})}{2\sinh(\frac{1}{2})}
+\frac{47}{24}, \quad x\in\mathbb{R}.
\end{array}
$$
The graph of $g_{1}(x)-g_{\mu}(x)$ can be seen in Fig.3 in \cite{lmt10}. Note that $\mu(g)=1$.

The inverse $v=A_{\mu}^{-1}w$ is given by
$$
\begin{array}{rl}
v(x)
&=(g*w)(x)=(g_{\mu}*w)(x)-(g_{1}*w)(x)+(1*w)(x)\\[5pt]
&=(\mu-\partial_{x}^{2})^{-1}w-(1-\partial_{x}^{2})^{-1}w+\int_{0}^{1}w(x)dx\\[5pt]
&=(\frac{x^{2}}{2}-\frac{x}{2}+\frac{25}{12})\mu(w)
+(x-\frac{1}{2})\int_{0}^{1}\int_{0}^{y}w(s)dsdy
+\int_{0}^{1}\int_{0}^{y}\int_{0}^{s}w(r)drdsdy\\[5pt]
&\quad-\int_{0}^{x}\int_{0}^{y}w(s)dsdy
-(1-\partial_{x}^{2})^{-1}w.
\end{array}
$$
Since $(\mu-\partial_{x}^{2})^{-1}$ and $(1-\partial_{x}^{2})^{-1}$ commute with
$\partial_{x}$, the following identity holds
$$
\begin{array}{rl}
A_{\mu}^{-1}\partial_{x}w
=\partial_{x}A_{\mu}^{-1}w,
\end{array}
$$
that is, $A_{\mu}^{-1}$ commutes with $\partial_{x}$.

For any $s\in \mathbb{R}$, $H^{s}(\mathbb{S})$ is defined by the Sobolev space of periodic functions
$$
H^{s}(\mathbb{S})
=\{v=\sum_{k}\widehat{v}(k)e^{2\pi ikx}:~
\|v\|_{H^{s}(\mathbb{S})}^{2}=\sum |\widehat{\Lambda^{s}v}(k)|^{2}<\infty\},
$$
where the pseudodifferential operator $\Lambda^{s}=(1-\partial_{x}^{2})^{\frac{s}{2}}$ is defined by
$$
\widehat{\Lambda^{s}v}(k)=(1+4\pi^{2}k^{2})^{\frac{s}{2}}\widehat{v}(k).
$$
We can check that $A_{\mu}=\mu-\partial_{x}^{2}+\partial_{x}^{4}$ is an isomorphism between $H^{s}(\mathbb{S})$ and $H^{s-4}(\mathbb{S})$.
Moreover, when $w\in H^{r+j-4}(\mathbb{S)}$ for $j=0,1,2,3$, we have
$A_{\mu}^{-1}\partial_{x}^{j}w\in H^{r}(\mathbb{S)}$ with
$$
\begin{array}{rl}
\|A_{\mu}^{-1}\partial_{x}^{j}w\|_{H^{r}(\mathbb{S)}}^{2}
&=\sum_{k}(1+4\pi^{2}k^{2})^{r}|\widehat{A_{\mu}^{-1}\partial_{x}^{j}w}(k)|^{2}\\[5pt]
&=\sum_{k}(1+4\pi^{2}k^{2})^{r}|\frac{(2\pi ik)^{j}}{\delta_{0}(k)+(2\pi k)^{2}+(2\pi k)^{4}}\widehat{w}(k)|^{2}\\[5pt]
&=\sum_{k}(1+4\pi^{2}k^{2})^{r+j-4}(1+4\pi^{2}k^{2})^{4-j}|\frac{(2\pi ik)^{j}}{\delta_{0}(k)+(2\pi k)^{2}+(2\pi k)^{4}}|^{2}|\widehat{w}(k)|^{2}\\[5pt]
&\leq 2^{4-j}\sum_{k}(1+4\pi^{2}k^{2})^{r+j-4}|\widehat{w}(k)|^{2}\\[5pt]
&=2^{4-j}\|w\|_{H^{r+j-4}(\mathbb{S)}}^{2}.
\end{array}
\eqno(2.1)
$$

\subsection{Local well-posedness}

The initial-value problem associated to Eq.
(1.4) can be rewritten in the following form:
$$
\begin{array}{l}
\left\{\begin{array}{l}
\mu(u_{t})-u_{txx}+u_{txxxx}+2\mu(u)u_{x}-2u_{x}u_{xx}-uu_{xxx}+2u_{x}u_{xxxx}+uu_{xxxxx}=0, \quad t>0, ~x\in \mathbb{R}, \\[3pt]
u(t,x+1)=u(t,x), \quad t\geq0, ~x\in \mathbb{R},\\[3pt]
u(0,x)=u_{0}(x), \quad x\in \mathbb{R}.
\end{array}
\right.
\end{array}
\eqno(2.2)
$$
or, equivalently,
$$
\begin{array}{l}
\left\{\begin{array}{l}
u_{t}+uu_{x}
 +\partial_{x}A_{\mu}^{-1}\left(2\mu(u)u+\frac{1}{2}u_{x}^{2}-3u_{x}u_{xxx}-\frac{7}{2}u_{xx}^{2}\right)=0,\quad t>0, ~x\in \mathbb{R},\\[3pt]
u(t,x+1)=u(t,x), \quad t\geq0, ~x\in \mathbb{R},\\[3pt]
u(0,x)=u_{0}(x), \quad x\in \mathbb{R},
\end{array}
\right.
\end{array}
\eqno(2.3)
$$

On the other hand, integrating both sides of Eq. (2.3) over $\mathbb{S}$ with respect to $x$, we obtain
$$
\frac{d}{dt}\mu(u)=0.
$$
Then it follows that
$$
\mu(u)=\mu(u_{0}):=\mu_{0}.
$$
Thus, Eq. (2.3) can be rewritten as
$$
\begin{array}{l}
\left\{\begin{array}{l}
u_{t}+uu_{x}
 +\partial_{x}A_{\mu}^{-1}\left(2\mu_{0}u+\frac{1}{2}u_{x}^{2}-3u_{x}u_{xxx}-\frac{7}{2}u_{xx}^{2}\right)=0, \quad t>0, ~x\in \mathbb{S},\\[3pt]
u(0,x)=u_{0}(x), \quad x\in \mathbb{S}.
\end{array}
\right.
\end{array}
\eqno(2.4)
$$

Applying the Kato's theorem \cite{k75}, one may follow the similar argument as in \cite{ly14} to obtain the following local well-posedness result for Eq. (2.4).\\

\noindent\textbf{Theorem 2.1.}  Given $u_{0}\in H^{s}(\mathbb{S}), s>\frac{7}{2}$,
there exist a maximal $T=T(u_{0})>0$, and a unique solution
$u$ to Eq. (2.4) such that
$$
u=u(\cdot,u_{0})\in C([0,T);H^{s}(\mathbb{S}))\cap C^{1}([0,T);H^{s-1}(\mathbb{S})).
$$
Moreover, the solution depends continuously on the initial data, and $T$ is independent of $s$.\\

\noindent\textbf{Lemma 2.1.}
(See \cite{ca00})
Assume $f(x)\in H^{1}(\mathbb{S})$ satisfies that $\int_{\mathbb{S}}f(x)dx=\frac{a_{0}}{2}$.
Then, for any $\varepsilon>0$, we have
$$
\max_{x\in\mathbb{S}}f^{2}(x)
\leq \frac{\varepsilon+2}{24}\int_{\mathbb{S}}f_{x}^{2}(x)dx+\frac{\varepsilon+2}{4\varepsilon}a_{0}^{2}.
$$

\noindent\textbf{Corollary 2.1.}
For $f(x)\in H^{1}(\mathbb{S})$, if $\int_{\mathbb{S}}f(x)dx=0$,
then we have
$$
\max_{x\in\mathbb{S}}f^{2}(x)
\leq \frac{1}{12}\int_{\mathbb{S}}f_{x}^{2}(x)dx.
$$

\noindent\textbf{Lemma 2.2.}
Let $u_{0}\in H^{s}(\mathbb{S}), s>\frac{7}{2}$,
and let $T$ be the maximal existence time of the solution $u$ to Eq. (2.4) with the initial data $u_{0}$.
Then we have
$$
\begin{array}{l}
\int_{\mathbb{S}}(u_{x}^{2}+u_{xx}^{2})dx
    =\int_{\mathbb{S}}(u_{0,x}^{2}+u_{0,xx}^{2})dx:=\mu_{1}^{2}, \quad \forall ~t\in [0,T).
\end{array}
\eqno(2.5)
$$
Moreover, we have
$$
\|u_{x}(t,\cdot)\|_{L^{\infty}(\mathbb{S})}
\leq \frac{\sqrt{3}}{6}\mu_{1}
$$
and
$$
\|u(t,\cdot)\|_{L^{\infty}(\mathbb{S})}
\leq |\mu_{0}|+\frac{1}{12}\mu_{1}.
$$

\noindent\textbf{Proof.}
A direct computation gives
$$
\frac{d}{dt}\int_{\mathbb{S}}(u_{x}^{2}+u_{xx}^{2})dx=0,
\quad\mbox{which implies (2.5)}.
$$
Since $u(t, \cdot)\in H^{s}(\mathbb{S})\subset C^{3}(\mathbb{S})$ for $s>\frac{7}{2}$, and $\int_{\mathbb{S}}u_{x}dx=0$, Corollary 2.1 and $(2.5)$ implies that
$$
\max_{x\in\mathbb{S}}u_{x}^{2}(t,x)
\leq \frac{1}{12}\int_{\mathbb{S}}u_{xx}^{2}(t,x)dx
\leq\frac{1}{12}\int_{\mathbb{S}}(u_{0,x}^{2}+u_{0,xx}^{2})dx
=\frac{1}{12}\mu_{1}^{2}.
$$
It then follows that
$$
\|u_{x}(t,\cdot)\|_{L^{\infty}(\mathbb{S})}
\leq \frac{\sqrt{3}}{6}\mu_{1}.
$$
Note that
$$
\int_{\mathbb{S}}(u(t,x)-\mu_{0})=0.
$$
By Corollary 2.1, we have
$$
\max_{x\in\mathbb{S}}(u(t,x)-\mu_{0})^{2}
\leq \frac{1}{12}\int_{\mathbb{S}}u_{x}^{2}(t,x)dx
\leq \frac{1}{12}\|u_{x}(t,\cdot)\|_{L^{\infty}(\mathbb{S})}^{2},
$$
which implies that
$$
\|u(t,\cdot)\|_{L^{\infty}(\mathbb{S})}-|\mu_{0}|
\leq \|u(t,\cdot)-\mu_{0}\|_{L^{\infty}(\mathbb{S})}
\leq \frac{1}{12}\mu_{1}.
$$
Hence, we get
$$
\|u(t,\cdot)\|_{L^{\infty}(\mathbb{S})}
\leq |\mu_{0}|+\frac{1}{12}\mu_{1}.
$$
This completes the proof of the lemma.  \hfill $\Box$

\section{Global existence of strong solution}

In this section, we present the global existence of strong solution to Eq. (2.4). Firstly, we will give some useful lemmas.\\

\noindent\textbf{Lemma 3.1.} (see \cite{kap88}) If $r>0$, then $H^{r}(\mathbb{S})\cap L^{\infty}(\mathbb{S})$
is an algebra. Moreover,
$$
\|fg\|_{H^{r}(\mathbb{S})}
\leq c_{r}(\|f\|_{L^{\infty}(\mathbb{S})}\|g\|_{H^{r}(\mathbb{S})}
 +\|f\|_{H^{r}(\mathbb{S})}\|g\|_{L^{\infty}(\mathbb{S})}),
$$
where $c_{r}$ is a positive constant depending only on $r$.\\

\noindent\textbf{Lemma 3.2.} (see \cite{kap88}) If $r>0$, then
$$
\left\|[\Lambda^{r}, f]g\right\|_{L^{2}(\mathbb{S})}
\leq c_{r}(\|\partial_{x}f\|_{L^{\infty}(\mathbb{S})}\|\Lambda^{r-1}g\|_{L^{2}(\mathbb{S})}
 +\|\Lambda^{r}f\|_{L^{2}(\mathbb{S})}\|g\|_{L^{\infty}(\mathbb{S})}),
$$
where $\Lambda^{r}=(1-\partial_{x}^{2})^{r/2}$ and $c_{r}$ is a positive constant depending only on $r$.\\

\noindent\textbf{Lemma 3.3.} (see \cite{tay14,esko14}) If $f\in H^{s}(\mathbb{S})$ with $s>\frac{3}{2}$, then there exists a constant $c>0$ such that for any $g\in L^{2}(\mathbb{S})$ we have
$$
\|[J_{\varepsilon}, f]\partial_{x}g\|_{L^{2}(\mathbb{S})}
\leq c\|f\|_{C^{1}(\mathbb{S})}\|g\|_{L^{2}(\mathbb{S})},
$$
in which for each $\varepsilon\in(0,1]$, the operator $J_{\varepsilon}$ is the Friedrichs mollifier defined by
$$
\begin{array}{l}
J_{\varepsilon}f(x)=j_{\varepsilon}*f(x),
\end{array}
\eqno(3.1)
$$
where $j_{\varepsilon}(x)=\frac{1}{\varepsilon}j(\frac{x}{\varepsilon})$ and $j(x)$ is a nonnegative, even, smooth bump function supported in the interval $(-\frac{1}{2}, \frac{1}{2})$ such that $\int_{\mathbb{R}}j(x)dx=1$.
For any $f\in H^{s}(\mathbb{S})$ with $s\geq0$, we have $J_{\varepsilon}f\rightarrow f$ in $H^{s}(\mathbb{S})$ as $\varepsilon\rightarrow 0$. Moreover, for any $p\geq1$, the Young's inequality
$\|j_{\varepsilon}*f\|_{L^{p}(\mathbb{S})}\leq \|j_{\varepsilon}\|_{L^{1}(\mathbb{R})}\|f\|_{L^{p}(\mathbb{S})}=\|f\|_{L^{p}(\mathbb{S})}$ holds since $j(x)$ is supported in the interval $(-\frac{1}{2}, \frac{1}{2})$.\\

Now we give the following theorem, which is a sufficient condition of global existence of solution to Eq. (2.4).\\

\noindent\textbf{Theorem 3.1.} Let $u_{0}\in H^{s}(\mathbb{S}), s>\frac{7}{2}$,
and let $T$ be the maximal existence time of the solution $u$ to Eq. (2.4) with the initial data $u_{0}$. If there exists $K>0$ such that
$$
\|u_{xxx}(t,\cdot)\|_{L^{\infty}(\mathbb{S})}\leq K, \quad t\in [0,T),
$$
then the $H^{s}(\mathbb{S})$-norm of $u(t,\cdot)$ does not blow up on $[0,T)$.\\

\noindent\textbf{Proof.}
Note that the product $uu_{x}$ only has the regularity of $H^{s-1}(\mathbb{S})$ when $u\in H^{s}(\mathbb{S})$. To deal with this problem, we will consider the following modified equation
$$
\begin{array}{l}
(J_{\varepsilon}u)_{t}+J_{\varepsilon}(uu_{x})
 +\partial_{x}A_{\mu}^{-4}\left(2\mu_{0}J_{\varepsilon}u
 +\frac{1}{2}J_{\varepsilon}(u_{x}^{2})
 -3J_{\varepsilon}(u_{x}u_{xxx})-\frac{7}{2}J_{\varepsilon}(u_{xx}^{2})\right)=0,
\end{array}
\eqno(3.2)
$$
where $J_{\varepsilon}$ is defined in (3.1).

Applying the operator $\Lambda^{s}=(1-\partial_{x}^{2})^{s/2}$ to Eq. (3.2), then multiplying the resulting equation
by $\Lambda^{s}J_{\varepsilon}u$ and integrating with respect to $x\in \mathbb{S}$, we obtain
$$
\begin{array}{rl}
\frac{1}{2}\frac{d}{dt}\|J_{\varepsilon}u\|_{H^{s}(\mathbb{S})}^{2}
&=-\left(\Lambda^{s}J_{\varepsilon}(uu_{x}), \Lambda^{s}J_{\varepsilon}u\right)\\[3pt]
&\quad -\left(\Lambda^{s}J_{\varepsilon}u,\partial_{x}\Lambda^{s}A_{\mu}^{-4}\left(2\mu_{0}J_{\varepsilon}u
     +\frac{1}{2}J_{\varepsilon}(u_{x}^{2})
     -3J_{\varepsilon}(u_{x}u_{xxx})-\frac{7}{2}J_{\varepsilon}(u_{xx}^{2})\right)\right).
\end{array}
\eqno(3.3)
$$
In what follows next we use the fact that $\Lambda^{s}$ and $J_{\varepsilon}$ commute and that
$J_{\varepsilon}$ satisfies the properties
$$
(J_{\varepsilon}f, g)=(f, J_{\varepsilon}g)
\quad \mbox{and}\quad
\|J_{\varepsilon}u\|_{H^{s}(\mathbb{S})}\leq \|u\|_{H^{s}(\mathbb{S})}.
$$
Let us estimate the first term of the right hand side of (3.3).
$$
\begin{array}{rl}
&\left|\left(\Lambda^{s}J_{\varepsilon}(uu_{x}), \Lambda^{s}J_{\varepsilon}u\right)\right|\\[3pt]
&=\left|(\Lambda^{s}(uu_{x}),J_{\varepsilon}\Lambda^{s}J_{\varepsilon}u)\right|\\[3pt]
&=\left|([\Lambda^{s}, u]u_{x},J_{\varepsilon}\Lambda^{s}J_{\varepsilon}u)
  +( u\Lambda^{s}u_{x},J_{\varepsilon}\Lambda^{s}J_{\varepsilon}u)\right|\\[3pt]
&=\left|([\Lambda^{s}, u]u_{x},J_{\varepsilon}\Lambda^{s}J_{\varepsilon}u)
  +( J_{\varepsilon}u\partial_{x}\Lambda^{s}u,\Lambda^{s}J_{\varepsilon}u)\right|\\[3pt]
&=\left|([\Lambda^{s}, u]u_{x},J_{\varepsilon}\Lambda^{s}J_{\varepsilon}u)
  +([J_{\varepsilon},u]\partial_{x}\Lambda^{s}u,\Lambda^{s}J_{\varepsilon}u)
  +(uJ_{\varepsilon}\partial_{x}\Lambda^{s}u,\Lambda^{s}J_{\varepsilon}u)\right|\\[3pt]
&\leq\|[\Lambda^{s}, u]u_{x}\|_{L^{2}(\mathbb{S})}
  \|J_{\varepsilon}\Lambda^{s}J_{\varepsilon}u\|_{L^{2}(\mathbb{S})}
  +\|[J_{\varepsilon},u]\partial_{x}\Lambda^{s}u\|_{L^{2}(\mathbb{S})}
  \|\Lambda^{s}J_{\varepsilon}u\|_{L^{2}(\mathbb{S})}\\[3pt]
&\quad+\frac{1}{2}\left|(u_{x}\Lambda^{s}J_{\varepsilon}u,\Lambda^{s}J_{\varepsilon}u)\right|\\[3pt]
&\lesssim \|u\|_{C^{1}(\mathbb{S})}\|u\|_{H^{s}(\mathbb{S})}^{2},
\end{array}
\eqno(3.4)
$$
where we have used Lemma 3.2 with $r=s$ and Lemma 3.3. Here and in what follows, we use $``\lesssim"$ to denote inequality up to a positive constant. Furthermore, we estimate the second term of the right hand side of $(3.3)$ in the following way
$$
\begin{array}{rl}
&\left|\left(\Lambda^{s}J_{\varepsilon}u,\partial_{x}\Lambda^{s}A_{\mu}^{-4}\left(2\mu_{0}J_{\varepsilon}u
     +\frac{1}{2}J_{\varepsilon}(u_{x}^{2})
     -3J_{\varepsilon}(u_{x}u_{xxx})-\frac{7}{2}J_{\varepsilon}(u_{xx}^{2})\right)\right)\right|\\[3pt]
&\leq\left\|\partial_{x}A_{\mu}^{-4}\left(2\mu_{0}J_{\varepsilon}u
     +\frac{1}{2}J_{\varepsilon}(u_{x}^{2})
     -3J_{\varepsilon}(u_{x}u_{xxx})-\frac{7}{2}J_{\varepsilon}(u_{xx}^{2})\right)\right\|_{H^{s}(\mathbb{S})}\|u\|_{H^{s}(\mathbb{S})}\\[3pt]
&\lesssim\|2\mu_{0}J_{\varepsilon}u+\frac{1}{2}J_{\varepsilon}(u_{x}^{2})
     -3J_{\varepsilon}(u_{x}u_{xxx})-\frac{7}{2}J_{\varepsilon}(u_{xx}^{2})\|_{H^{s-3}(\mathbb{S})}\|u\|_{H^{s}(\mathbb{S})}\\[3pt]
&\lesssim (|\mu_{0}|\|u\|_{H^{s-3}(\mathbb{S})}
  +\|u_{x}\|_{L^{\infty}(\mathbb{S})}\|u\|_{H^{s-2}(\mathbb{S})}
  +\|u_{x}\|_{L^{\infty}(\mathbb{S})}\|u\|_{H^{s}(\mathbb{S})}
  +\|u_{xxx}\|_{L^{\infty}(\mathbb{S})}\|u\|_{H^{s-2}(\mathbb{S})}\\[3pt]
&\qquad  +\|u_{xx}\|_{L^{\infty}(\mathbb{S})}\|u_{xx}\|_{H^{s-3}(\mathbb{S})})\|u\|_{H^{s}(\mathbb{S})}\\[3pt]
&\leq \left(|\mu_{0}|
  +\|u_{x}\|_{L^{\infty}(\mathbb{S})}
  +\|u_{xx}\|_{L^{\infty}(\mathbb{S})}
  +\|u_{xxx}\|_{L^{\infty}(\mathbb{S})}\right)\|u\|_{H^{s}(\mathbb{S})}^{2},
\end{array}
$$
where we have used Lemma 3.1 and (2.1).

Since $u(t, \cdot)\in H^{s}(\mathbb{S})\subset C^{3}(\mathbb{S})$ for $s>\frac{7}{2}$, and $\int_{\mathbb{S}}u_{xx}dx=0$, Corollary 2.1 implies that
$$
\|u_{xx}(t,\cdot)\|_{L^{\infty}(\mathbb{S})}
\leq \frac{\sqrt{3}}{6}\|u_{xxx}(t,\cdot)\|_{L^{2}(\mathbb{S})}
\leq \frac{\sqrt{3}}{6}\|u_{xxx}(t,\cdot)\|_{L^{\infty}(\mathbb{S})}, \quad t\in [0,T).
$$
Thus,
$$
\begin{array}{rl}
&\left|\left(\Lambda^{s}J_{\varepsilon}u,\partial_{x}\Lambda^{s}A_{\mu}^{-4}\left(2\mu_{0}J_{\varepsilon}u
     +\frac{1}{2}J_{\varepsilon}(u_{x}^{2})
     -3J_{\varepsilon}(u_{x}u_{xxx})-\frac{7}{2}J_{\varepsilon}(u_{xx}^{2})\right)\right)\right|\\[3pt]
&\lesssim \left(|\mu_{0}|
  +\|u_{x}\|_{L^{\infty}(\mathbb{S})}
  +\|u_{xxx}\|_{L^{\infty}(\mathbb{S})}\right)\|u\|_{H^{s}(\mathbb{S})}^{2},
\end{array}
\eqno(3.5)
$$
Combining $(3.4)$ and $(3.5)$
and using Lemma 2.2, we have
$$
\begin{array}{rl}
\frac{1}{2}\frac{d}{dt}\|J_{\varepsilon}u\|_{H^{s}(\mathbb{S})}^{2}
&\lesssim \left(|\mu_{0}|
  +\|u\|_{L^{\infty}(\mathbb{S})}
  +\|u_{x}\|_{L^{\infty}(\mathbb{S})}
  +\|u_{xxx}\|_{L^{\infty}(\mathbb{S})}\right)\|u\|_{H^{s}(\mathbb{S})}^{2}\\[5pt]
&\lesssim (|\mu_{0}|+\mu_{1}+\|u_{xxx}\|_{L^{\infty}(\mathbb{S})})\|u\|_{H^{s}(\mathbb{S})}^{2}.
\end{array}
\eqno(3.6)
$$
Letting $\varepsilon\rightarrow 0$, we get
$$
\begin{array}{l}
\frac{1}{2}\frac{d}{dt}\|u\|_{H^{s}(\mathbb{S})}^{2}
\leq c(1+\|u_{xxx}\|_{L^{\infty}(\mathbb{S})})\|u\|_{H^{s}(\mathbb{S})}^{2},
\end{array}
$$
where $c$ is a constant depending on $s$ and $u_{0}$.
An application of Gronwall's inequality and the assumption of the theorem yield
$$
\|u\|_{H^{s}(\mathbb{S})}^{2}
\leq e^{2c(1+K)t}\|u_{0}\|_{H^{s}(\mathbb{S})}^{2},
$$
which completes the proof of the theorem.  \hfill $\Box$ \\

\noindent\textbf{Theorem 3.2.}
Let $u_{0}\in H^{s}(\mathbb{S}), s>\frac{7}{2}$. Then the corresponding strong solution $u$ of the initial value $u_{0}$ exists globally in time.\\

\noindent\textbf{Proof.}
By using the local well-posedness theorem and a density argument, it suffices
to show the theorem for $s\geq 5$. Assume that $u_{0}\in H^{s}(\mathbb{S})$, $s\geq 5$. Let $u$ be the corresponding solution of Eq. (2.4) on $[0, T)\times \mathbb{S}$, which is guaranteed by Theorem 2.1. Multiplying Eq. (1.4) by $m$ and integrating over $\mathbb{S}$ with respect to $x$ yield
$$
\frac{1}{2}\frac{d}{dt}\int_{\mathbb{S}}m^{2}dx
=\int_{\mathbb{S}}m(-2mu_{x}-um_{x})dx
=-3\int_{\mathbb{S}}u_{x}m^{2}dx.
\eqno(3.7)
$$
Note that in Lemma 2.2 we have
$$
\|u_{x}(t,\cdot)\|_{L^{\infty}(\mathbb{S})}
\leq \frac{\sqrt{3}}{6}\mu_{1}.
$$
Then
$$
\frac{d}{dt}\int_{\mathbb{S}}m^{2}dx
\leq \sqrt{3}\mu_{1}\int_{\mathbb{S}}m^{2}dx.
$$
By Gronwall's inequality, we have
$$
\int_{\mathbb{S}}m^{2}dx\leq e^{\sqrt{3}\mu_{1}t}\int_{\mathbb{S}}m_{0}^{2}dx.
$$
Note that
$$
\int_{\mathbb{S}}m^{2}
=\mu(u)^{2}+\int_{\mathbb{S}}u_{xx}^{2}+2\int_{\mathbb{S}}u_{xxx}^{2}+\int_{\mathbb{S}}u_{xxxx}^{2}\geq\|u_{xxxx}\|_{L^{2}(\mathbb{S})}^{2}.
$$
Since $u\in H^{5}(\mathbb{S})\subset C^{4}(\mathbb{S})$ and $\int_{\mathbb{S}}u_{xxx}dx=0$,
Corollary 2.1 implies that
$$
\begin{array}{l}
\|u_{xxx}\|_{L^{\infty}(\mathbb{S})}
\leq \frac{\sqrt{3}}{6}\|u_{xxxx}\|_{L^{2}(\mathbb{S})}
\leq \frac{\sqrt{3}}{6}\|m\|_{L^{2}(\mathbb{S})}
\leq \frac{\sqrt{3}}{6}e^{\frac{\sqrt{3}}{2}\mu_{1}t}\|m_{0}\|_{L^{2}(\mathbb{S})}.
\end{array}
$$
Theorem 3.1 ensures that the solution $u$ does not blow up in finite time, that is, $T=\infty$. This completes the proof
of Theorem 3.2. \hfill $\Box$

\section{Non-uniform dependence on initial data}

In this section, we will first give an estimate of the solution size in time interval
$[0, T_{0}]$ for any fixed $T_{0}>0$,
and then we show that, for any $s\geq 4$, the data-to-solution map is H\"{o}lder continuous from any bounded subset of $H^{s}(\mathbb{S})$ into $C([0, T_{0}]; H^{r}(\mathbb{S}))$ with $0\leq r<s$, but is not uniformly continuous from any bounded subset of $H^{s}(\mathbb{S})$ into $C([0, T_{0}]; H^{s}(\mathbb{S}))$.\\

Similar as the proofs of in \cite{wang17}, we can obtain the following Lemma 4.1 and Theorem 4.1.\\

\noindent\textbf{Lemma 4.1.} Let $u$ be the solution of Eq. (2.4) with initial data $u_{0}\in H^{s}(\mathbb{S}), s\geq 4$. Then, for any fixed $T_{0}>0$, we have
$$
\begin{array}{l}
\|u(t)\|_{H^{s}(\mathbb{S})}\leq
e^{cT_{0}}\|u_{0}\|_{H^{s}(\mathbb{S})},
\quad t\in[0, T_{0}],
\end{array}
\eqno(4.1)
$$
where $c=c(s, T_{0}, u_{0})$ is a constant depending on $s$, $T_{0}$ and $\|u_{0}\|_{H^{4}(\mathbb{S})}$.\\

\noindent\textbf{Theorem 4.1.} Assume $s\geq4$ and $0\leq r<s$. Then the solution map
for Eq. (2.4) is H\"{o}lder continuous with exponent
$$
\begin{array}{rl}
\alpha=
\left\{\begin{array}{l}
1, \quad \mbox{if}~0\leq r\leq s-1, \\[3pt]
s-r, \quad \mbox{if}~s-1< r< s
\end{array}
\right.
\end{array}
$$
as a map from $B(0,h)$ with $H^{r}(\mathbb{S})$-norm to $C([0,T_{0}]; H^{r}(\mathbb{S}))$
for any fixed $T_{0}>0$. More precisely, we have
$$
\|u(t)-w(t)\|_{C([0,T_{0}]; H^{r}(\mathbb{S}))}
\leq c\|u(0)-w(0)\|_{H^{r}(\mathbb{S})}^{\alpha},
$$
for all $u(0), w(0)\in B(0,h):=\{u\in H^{s}(\mathbb{S}): \|u\|_{H^{s}(\mathbb{S})}\leq h\}$
and $u(t), w(t)$ the solutions corresponding to the initial data $u(0), w(0)$,
respectively. The constant $c$ depends on $s, r, T_{0}$ and $h$.\\

Next, we prove that the data-to-solution map is not uniformly continuous. Firstly, we will recall some useful lemmas.\\

\noindent\textbf{Lemma 4.2.} (see \cite{himkm10}) Let $\sigma, \alpha\in \mathbb{R}$. If $n\in \mathbb{Z}^{+}$ and
$n\gg 1$, then
$$
\|\cos(nx-\alpha)\|_{H^{\sigma}(\mathbb{S})}
\approx n^{\sigma}.
$$
Relation is also true if $\cos(nx-\alpha)$ is replaced by $\sin(nx-\alpha)$.\\

\noindent\textbf{Lemma 4.3.} (see \cite{taylor03}) If $s>\frac{3}{2}$ and $0\leq \sigma+1\leq s$, then there exists a constant $c>0$ such that
$$
\|[\Lambda^{\sigma}\partial_{x}, f]v\|_{L^{2}(\mathbb{S})}
\leq c\|f\|_{H^{s}(\mathbb{S})}\|v\|_{H^{\sigma}(\mathbb{S})}.
$$

\noindent\textbf{Lemma 4.4.} (see \cite{himkm10}) If $r>\frac{1}{2}$, then there exists a constant $c_{r}>0$ depending only on $r$ such that
$$
\|fg\|_{H^{r-1}(\mathbb{S})}\leq c_{r}\|f\|_{H^{r}(\mathbb{S})}\|g\|_{H^{r-1}(\mathbb{S})}.
$$

\subsection{Approximate solutions}

The approximate solutions are of the form
$$
\begin{array}{l}
u^{\omega, n}(t, x)=\omega n^{-1}+n^{-s}\cos(nx-\omega t),
\end{array}
$$
where $\omega$ is in a bounded subset of $\mathbb{R}$ and $n\in \mathbb{Z}^{+}$.
Now we compute the error of the approximate solutions.

Note that
$$
\begin{array}{l}
\partial_{t}u^{\omega, n}=\omega n^{-s}\sin(nx-\omega t)
\end{array}
$$
and
$$
\begin{array}{l}
\partial_{x}u^{\omega, n}=-n^{-s+1}\sin(nx-\omega t).
\end{array}
$$
Since $A_{\mu}^{-1}$ commutes with $\partial_{x}$, we have
$$
\begin{array}{l}
F_{1}:=\partial_{t}u^{\omega, n}+u^{\omega, n}\partial_{x}u^{\omega, n}
=-\frac{1}{2}n^{-2s+1}\sin(2nx-2\omega t),\\[5pt]
F_{2}:=2\mu(u^{\omega, n})\partial_{x}A_{\mu}^{-1}u^{\omega, n}
=-2n^{-s+1}\mu(u^{\omega, n})A_{\mu}^{-1}\sin(nx-\omega t)\\[5pt]
\quad~=-2n^{-s+1}[\omega n^{-1}+n^{-s-1}(\sin(n-\omega t)+\sin(\omega t))]
A_{\mu}^{-1}\sin(nx-\omega t),\\[5pt]
F_{3}:=\frac{1}{2}\partial_{x}A_{\mu}^{-1}(u_{x}^{\omega, n})^{2}
=\frac{1}{2}n^{-2s+3}A_{\mu}^{-1}\sin(2nx-2\omega t), \\[5pt]
F_{4}:=-3\partial_{x}A_{\mu}^{-1}u_{x}^{\omega, n}u_{xxx}^{\omega, n}
=3n^{-2s+5}A_{\mu}^{-1}\sin(2nx-2\omega t), \\[5pt]
F_{5}:=\frac{7}{2}\partial_{x}A_{\mu}^{-1}(u_{xx}^{\omega, n})^{2}
=-\frac{7}{2}n^{-2s+5}A_{\mu}^{-1}\sin(2nx-2\omega t).
\end{array}
$$
By (2.1) and Lemma 4.2, for $n\gg 1$, we have
$$
\begin{array}{l}
\|F_{1}(t, \cdot)\|_{H^{\sigma}(\mathbb{S})}
=\frac{1}{2}n^{-2s+1}\|\sin(2nx-2\omega t)\|_{H^{\sigma}(\mathbb{S})}
\lesssim n^{-2s+1+\sigma},\\[5pt]
\|F_{2}(t, \cdot)\|_{H^{\sigma}(\mathbb{S})}
=\|-2n^{-s+1}[\omega n^{-1}+n^{-s-1}(\sin(n-\omega t)+\sin(\omega t))]
A_{\mu}^{-1}\sin(nx-\omega t)\|_{H^{\sigma}(\mathbb{S})} \\[5pt]
\qquad\qquad\qquad\lesssim (n^{-s}+n^{-2s})\|A_{\mu}^{-1}\sin(nx-\omega t)\|_{H^{\sigma}(\mathbb{S})}
\lesssim (n^{-s}+n^{-2s})\|\sin(nx-\omega t)\|_{H^{\sigma-4}(\mathbb{S})} \\[5pt]
\qquad\qquad\qquad\lesssim n^{-s-4+\sigma}+n^{-2s-4+\sigma},\\[5pt]
\|F_{3}(t, \cdot)\|_{H^{\sigma}(\mathbb{S})}
=\frac{1}{2}n^{-2s+3}\|A_{\mu}^{-1}\sin(2nx-2\omega t)\|_{H^{\sigma}(\mathbb{S})}
\lesssim n^{-2s+3}\|\sin(2nx-2\omega t)\|_{H^{\sigma-4}(\mathbb{S})}\\[5pt]
\qquad\qquad\qquad~\lesssim n^{-2s-1+\sigma}, \\[5pt]
\|F_{4}(t, \cdot)\|_{H^{\sigma}(\mathbb{S})}
=3n^{-2s+5}\|A_{\mu}^{-1}\sin(2nx-2\omega t)\|_{H^{\sigma}(\mathbb{S})}
\lesssim n^{-2s+5}\|\sin(2nx-2\omega t)\|_{H^{\sigma-4}(\mathbb{S})}\\[5pt]
\qquad\qquad\qquad~\lesssim n^{-2s+1+\sigma}, \\[5pt]
\|F_{5}(t, \cdot)\|_{H^{\sigma}(\mathbb{S})}
=\frac{7}{2}n^{-2s+5}\|A_{\mu}^{-1}\sin(2nx-2\omega t)\|_{H^{\sigma}(\mathbb{S})}
\lesssim n^{-2s+5}\|\sin(2nx-2\omega t)\|_{H^{\sigma-4}(\mathbb{S})}\\[5pt]
\qquad\qquad\qquad~\lesssim n^{-2s+1+\sigma}.
\end{array}
$$
Thus, for the error $F:=\sum_{i=1}^{5}F_{i}$ of the approximate solution, we have the following estimate.\\

\noindent\textbf{Lemma 4.5.} If $\omega$ is bounded, then for $n\gg 1$, we have
$$
\|F(t, \cdot)\|_{H^{\sigma}(\mathbb{S})}
\lesssim n^{-2s+1+\sigma}+n^{-s-4+\sigma}+n^{-2s-4+\sigma}
+n^{-2s-1+\sigma}.
$$
In particular, if $s>\frac{1+\sigma}{2}$, then
$$
\|F(t, \cdot)\|_{H^{\sigma}(\mathbb{S})}
\lesssim n^{-r_{s}},
$$
where $r_{s}>0$ and
$$
\begin{array}{l}
r_{s}
=\left\{\begin{array}{l}
2s-1-\sigma, \quad \mbox{if}~\frac{1+\sigma}{2}<s\leq 5,\\[5pt]
s+4-\sigma, \quad \mbox{if}~s> 5.
\end{array}
\right.
\end{array}
$$

\subsection{Difference between approximate and actual solutions}

Let $u_{\omega, n}$ be the solution of Eq. (2.4) with initial data given by the approximate solution
$u^{\omega, n}$ evaluated at time zero. That is, $u_{\omega, n}$ solves the following Cauchy problem
$$
\begin{array}{l}
\left\{\begin{array}{l}
\partial_{t}u_{\omega, n}+u_{\omega, n}\partial_{x}u_{\omega, n}
 +\partial_{x}A_{\mu}^{-1}(2\mu(u_{\omega, n})u_{\omega, n}+\frac{1}{2}(\partial_{x}u_{\omega, n})^{2}-3\partial_{x}u_{\omega, n}\partial_{x}^{3}u_{\omega, n}\\[3pt]
\qquad\qquad -\frac{7}{2}(\partial_{x}^{2}u_{\omega, n})^{2})=0, \quad t>0, ~x\in \mathbb{S},\\[3pt]
u_{\omega, n}(0,x)=u^{\omega, n}(0,x)=\omega n^{-1}+n^{-s}\cos(nx), \quad x\in \mathbb{S}.
\end{array}
\right.
\end{array}
\eqno(4.2)
$$
Using Lemma 4.2, we obtain
$$
\|u_{\omega, n}(0,\cdot)\|_{H^{s}(\mathbb{S})}
=\|u^{\omega, n}(0,\cdot)\|_{H^{s}(\mathbb{S})}
=\|\omega n^{-1}+n^{-s}\cos(nx)\|_{H^{s}(\mathbb{S})}
\lesssim 1.
$$
By Theorem 2.1, we know that $u_{\omega, n}$ is the unique solution of (4.2) and exists globally in time. To estimate the difference between the approximate and actual solutions, we let
$v=u^{\omega, n}-u_{\omega, n}$, then for $t>0$ and $x\in \mathbb{S}$, $v$ satisfies
the following Cauchy problem
$$
\begin{array}{l}
\left\{\begin{array}{l}
\partial_{t}v=F-\frac{1}{2}\partial_{x}[(u^{\omega, n}+u_{\omega, n})v]
-2\mu(u_{\omega, n})\partial_{x}A_{\mu}^{-1}v
-2\mu(v)\partial_{x}A_{\mu}^{-1}u^{\omega, n}\\[5pt]
\quad\quad
-\frac{1}{2}\partial_{x}A_{\mu}^{-1}[\partial_{x}(u^{\omega, n}+u_{\omega, n})\partial_{x}v]
+3\partial_{x}A_{\mu}^{-1}(\partial_{x}u^{\omega, n}\partial_{x}^{3}v)
+3\partial_{x}A_{\mu}^{-1}(\partial_{x}^{3}u_{\omega, n}\partial_{x}v)\\[5pt]
\quad\quad
+\frac{7}{2}\partial_{x}A_{\mu}^{-1}[\partial_{x}^{2}(u^{\omega, n}+u_{\omega, n})\partial_{x}^{2}v], \quad t>0, ~x\in \mathbb{S},\\[3pt]
v(0,x)=0, \quad x\in \mathbb{S}.
\end{array}
\right.
\end{array}
\eqno(4.3)
$$

\noindent\textbf{Lemma 4.6.} If $n\gg 1$, $s\geq 4$ and $\frac{5}{2}<\sigma\leq s$, then
for any fixed $T_{0}>0$, we have
$$
\|v(t, \cdot)\|_{H^{\sigma}(\mathbb{S})}
\lesssim n^{-r_{s}}, \quad t\in[0, T_{0}].
$$

\noindent\textbf{Proof.} Applying $\Lambda^{\sigma}$ to both sides of (4.3), multiplying the resulting equation by $\Lambda^{\sigma}v$ and integrating it with respect to $x$, we obtain
$$
\begin{array}{rl}
&\frac{1}{2}\frac{d}{dt}\|v(t, \cdot)\|_{H^{\sigma}(\mathbb{S})}^{2}\\[5pt]
&=\int_{\mathbb{S}}\Lambda^{\sigma}F\cdot \Lambda^{\sigma}vdx
-\frac{1}{2}\int_{\mathbb{S}}\Lambda^{\sigma}\partial_{x}[(u^{\omega, n}+u_{\omega, n})v]\cdot \Lambda^{\sigma}vdx\\[5pt]
&\quad
-2\mu(u_{\omega, n})\int_{\mathbb{S}}\Lambda^{\sigma}\partial_{x}A_{\mu}^{-1}v\cdot \Lambda^{\sigma}vdx
-2\mu(v)\int_{\mathbb{S}}\Lambda^{\sigma}\partial_{x}A_{\mu}^{-1}u^{\omega, n}\cdot \Lambda^{\sigma}vdx\\[5pt]
&\quad
-\frac{1}{2}\int_{\mathbb{S}}\Lambda^{\sigma}\partial_{x}A_{\mu}^{-1}[\partial_{x}(u^{\omega, n}+u_{\omega, n})\partial_{x}v]\cdot \Lambda^{\sigma}vdx
+3\int_{\mathbb{S}}\Lambda^{\sigma}\partial_{x}A_{\mu}^{-1}(\partial_{x}u^{\omega, n}\partial_{x}^{3}v)\cdot \Lambda^{\sigma}vdx\\[5pt]
&\quad
+3\int_{\mathbb{S}}\Lambda^{\sigma}\partial_{x}A_{\mu}^{-1}(\partial_{x}^{3}u_{\omega, n}\partial_{x}v)\cdot \Lambda^{\sigma}vdx
+\frac{7}{2}\int_{\mathbb{S}}\Lambda^{\sigma}\partial_{x}A_{\mu}^{-1}[\partial_{x}^{2}(u^{\omega, n}+u_{\omega, n})\partial_{x}^{2}v]\cdot \Lambda^{\sigma}vdx\\[5pt]
&:=\sum_{i=1}^{8}G_{i}.
\end{array}
$$
By H\"{o}lder inequality, we know
$$
\begin{array}{rl}
|G_{1}|=|\int_{\mathbb{S}}\Lambda^{\sigma}F\cdot \Lambda^{\sigma}vdx|
\leq \|\Lambda^{\sigma}F\|_{L^{2}(\mathbb{S})}\|\Lambda^{\sigma}v\|_{L^{2}(\mathbb{S})}
= \|F\|_{H^{\sigma}(\mathbb{S})}\|v\|_{H^{\sigma}(\mathbb{S})}.
\end{array}
$$
For $G_{2}$, we have
$$
\begin{array}{rl}
|G_{2}|
&=|-\frac{1}{2}\int_{\mathbb{S}}\Lambda^{\sigma}\partial_{x}[(u^{\omega, n}+u_{\omega, n})v]\cdot \Lambda^{\sigma}vdx|\\[5pt]
&=|-\frac{1}{2}\int_{\mathbb{S}}[\Lambda^{\sigma}\partial_{x}, (u^{\omega, n}+u_{\omega, n})]v\cdot \Lambda^{\sigma}vdx
-\frac{1}{2}\int_{\mathbb{S}}(u^{\omega, n}+u_{\omega, n})\Lambda^{\sigma}\partial_{x}v\cdot \Lambda^{\sigma}vdx|\\[5pt]
&=|-\frac{1}{2}\int_{\mathbb{S}}[\Lambda^{\sigma}\partial_{x}, (u^{\omega, n}+u_{\omega, n})]v\cdot \Lambda^{\sigma}vdx
+\frac{1}{4}\int_{\mathbb{S}}\partial_{x}(u^{\omega, n}+u_{\omega, n})\cdot(\Lambda^{\sigma}v)^{2}dx|\\[5pt]
&\leq \frac{1}{2}\|[\Lambda^{\sigma}\partial_{x}, (u^{\omega, n}+u_{\omega, n})]v\|_{L^{2}(\mathbb{S})}
\|\Lambda^{\sigma}v\|_{L^{2}(\mathbb{S})}
+\frac{1}{4}\|\partial_{x}(u^{\omega, n}+u_{\omega, n})\|_{L^{\infty}(\mathbb{S})}
\|\Lambda^{\sigma}v\|_{L^{2}(\mathbb{S})}^{2}\\[5pt]
&\lesssim \|u^{\omega, n}+u_{\omega, n}\|_{H^{s}(\mathbb{S})}\|v\|_{H^{\sigma}(\mathbb{S})}^{2},
\end{array}
$$
where we have used integrating by parts, the Sobolev imbedding theorem and Lemma 4.3.

According to (2.1), we have
$$
\begin{array}{rl}
&|G_{3}|+|G_{4}|\\[5pt]
&=|2\mu(u_{\omega, n})\int_{\mathbb{S}}\Lambda^{\sigma}\partial_{x}A_{\mu}^{-1}v\cdot \Lambda^{\sigma}vdx|
+|2\mu(v)\int_{\mathbb{S}}\Lambda^{\sigma}\partial_{x}A_{\mu}^{-1}u^{\omega, n}\cdot \Lambda^{\sigma}vdx|\\[5pt]
&\leq 2\|u_{\omega, n}\|_{L^{2}(\mathbb{S})}
\|\Lambda^{\sigma}\partial_{x}A_{\mu}^{-1}v\|_{L^{2}(\mathbb{S})}
\|\Lambda^{\sigma}v\|_{L^{2}(\mathbb{S})}
+2\|v\|_{L^{2}(\mathbb{S})}\|\Lambda^{\sigma}\partial_{x}A_{\mu}^{-1}u^{\omega, n}\|_{L^{2}(\mathbb{S})}
\|\Lambda^{\sigma}v\|_{L^{2}(\mathbb{S})}\\[5pt]
&=2\|u_{\omega, n}\|_{L^{2}(\mathbb{S})}
\|\partial_{x}A_{\mu}^{-1}v\|_{H^{\sigma}(\mathbb{S})}
\|v\|_{H^{\sigma}(\mathbb{S})}
+2\|v\|_{L^{2}(\mathbb{S})}\|\partial_{x}A_{\mu}^{-1}u^{\omega, n}\|_{H^{\sigma}(\mathbb{S})}
\|v\|_{H^{\sigma}(\mathbb{S})}\\[5pt]
&\lesssim \|u_{\omega, n}\|_{L^{2}(\mathbb{S})}
\|v\|_{H^{\sigma-3}(\mathbb{S})}
\|v\|_{H^{\sigma}(\mathbb{S})}
+\|v\|_{L^{2}(\mathbb{S})}\|u^{\omega, n}\|_{H^{\sigma-3}(\mathbb{S})}
\|v\|_{H^{\sigma}(\mathbb{S})}\\[5pt]
&\lesssim (\|u_{\omega, n}\|_{H^{s}(\mathbb{S})}+\|u^{\omega, n}\|_{H^{s}(\mathbb{S})})\|v\|_{H^{\sigma}(\mathbb{S})}^{2}.
\end{array}
$$
Since $\frac{5}{2}<\sigma\leq s$, by (2.1) and Lemma 4.4, we get
$$
\begin{array}{rl}
&|G_{5}|+|G_{8}|\\[5pt]
&=|\frac{1}{2}\int_{\mathbb{S}}\Lambda^{\sigma}\partial_{x}A_{\mu}^{-1}[\partial_{x}(u^{\omega, n}+u_{\omega, n})\partial_{x}v]\cdot \Lambda^{\sigma}vdx|+
|\frac{7}{2}\int_{\mathbb{S}}\Lambda^{\sigma}\partial_{x}A_{\mu}^{-1}[\partial_{x}^{2}(u^{\omega, n}+u_{\omega, n})\partial_{x}^{2}v]\cdot \Lambda^{\sigma}vdx|\\[5pt]
&\leq \frac{1}{2}
\|\Lambda^{\sigma}\partial_{x}A_{\mu}^{-1}[\partial_{x}(u^{\omega, n}+u_{\omega, n})\partial_{x}v]\|_{L^{2}(\mathbb{S})}
\|\Lambda^{\sigma}v\|_{L^{2}(\mathbb{S})}\\[5pt]
&\quad+\frac{7}{2}\|\Lambda^{\sigma}\partial_{x}A_{\mu}^{-1}[\partial_{x}^{2}(u^{\omega, n}+u_{\omega, n})\partial_{x}^{2}v]\|_{L^{2}(\mathbb{S})}
\|\Lambda^{\sigma}v\|_{L^{2}(\mathbb{S})}\\[5pt]
&=\frac{1}{2}\|\partial_{x}A_{\mu}^{-1}[\partial_{x}(u^{\omega, n}+u_{\omega, n})\partial_{x}v]\|_{H^{\sigma}(\mathbb{S})}
\|v\|_{H^{\sigma}(\mathbb{S})}
+\frac{7}{2}\|\partial_{x}A_{\mu}^{-1}[\partial_{x}^{2}(u^{\omega, n}+u_{\omega, n})\partial_{x}^{2}v]\|_{H^{\sigma}(\mathbb{S})}
\|v\|_{H^{\sigma}(\mathbb{S})}\\[5pt]
&\lesssim
\|\partial_{x}(u^{\omega, n}+u_{\omega, n})\partial_{x}v\|_{H^{\sigma-3}(\mathbb{S})}
\|v\|_{H^{\sigma}(\mathbb{S})}
+\|\partial_{x}^{2}(u^{\omega, n}+u_{\omega, n})\partial_{x}^{2}v\|_{H^{\sigma-3}(\mathbb{S})}
\|v\|_{H^{\sigma}(\mathbb{S})}\\[5pt]
&\lesssim
\|\partial_{x}(u^{\omega, n}+u_{\omega, n})\|_{H^{\sigma-2}(\mathbb{S})}
\|\partial_{x}v\|_{H^{\sigma-3}(\mathbb{S})}
\|v\|_{H^{\sigma}(\mathbb{S})}\\[5pt]
&\quad+\|\partial_{x}^{2}(u^{\omega, n}+u_{\omega, n})\|_{H^{\sigma-2}(\mathbb{S})}
\|\partial_{x}^{2}v\|_{H^{\sigma-3}(\mathbb{S})}
\|v\|_{H^{\sigma}(\mathbb{S})}\\[5pt]
&\lesssim (\|u_{\omega, n}\|_{H^{s}(\mathbb{S})}+\|u^{\omega, n}\|_{H^{s}(\mathbb{S})})\|v\|_{H^{\sigma}(\mathbb{S})}^{2}
\end{array}
$$
and
$$
\begin{array}{rl}
&|G_{6}|+|G_{7}|\\[5pt]
&=|3\int_{\mathbb{S}}\Lambda^{\sigma}\partial_{x}A_{\mu}^{-1}(\partial_{x}u^{\omega, n}\partial_{x}^{3}v)\cdot \Lambda^{\sigma}vdx|
+|3\int_{\mathbb{S}}\Lambda^{\sigma}\partial_{x}A_{\mu}^{-1}(\partial_{x}^{3}u_{\omega, n}\partial_{x}v)\cdot \Lambda^{\sigma}vdx|\\[5pt]
&\leq
3\|\Lambda^{\sigma}\partial_{x}A_{\mu}^{-1}(\partial_{x}u^{\omega, n}\partial_{x}^{3}v)\|_{L^{2}(\mathbb{S})}
\|\Lambda^{\sigma}v\|_{L^{2}(\mathbb{S})}
+3\|\Lambda^{\sigma}\partial_{x}A_{\mu}^{-1}(\partial_{x}^{3}u_{\omega, n}\partial_{x}v)\|_{L^{2}(\mathbb{S})}
\|\Lambda^{\sigma}v\|_{L^{2}(\mathbb{S})}\\[5pt]
&=3\|\partial_{x}A_{\mu}^{-1}(\partial_{x}u^{\omega, n}\partial_{x}^{3}v)\|_{H^{\sigma}(\mathbb{S})}
\|v\|_{H^{\sigma}(\mathbb{S})}
+3\|\partial_{x}A_{\mu}^{-1}(\partial_{x}^{3}u_{\omega, n}\partial_{x}v)\|_{H^{\sigma}(\mathbb{S})}
\|v\|_{H^{\sigma}(\mathbb{S})}\\[5pt]
&\lesssim
\|\partial_{x}u^{\omega, n}\partial_{x}^{3}v\|_{H^{\sigma-3}(\mathbb{S})}
\|v\|_{H^{\sigma}(\mathbb{S})}
+\|\partial_{x}^{3}u_{\omega, n}\partial_{x}v\|_{H^{\sigma-3}(\mathbb{S})}
\|v\|_{H^{\sigma}(\mathbb{S})}\\[5pt]
&\lesssim
\|\partial_{x}u^{\omega, n}\|_{H^{\sigma-2}(\mathbb{S})}
\|\partial_{x}^{3}v\|_{H^{\sigma-3}(\mathbb{S})}
\|v\|_{H^{\sigma}(\mathbb{S})}
+\|\partial_{x}v\|_{H^{\sigma-2}(\mathbb{S})}
\|\partial_{x}^{3}u_{\omega, n}\|_{H^{\sigma-3}(\mathbb{S})}
\|v\|_{H^{\sigma}(\mathbb{S})}\\[5pt]
&\lesssim (\|u_{\omega, n}\|_{H^{s}(\mathbb{S})}+\|u^{\omega, n}\|_{H^{s}(\mathbb{S})})\|v\|_{H^{\sigma}(\mathbb{S})}^{2}.
\end{array}
$$
Thus,
$$
\begin{array}{rl}
\frac{1}{2}\frac{d}{dt}\|v(t, \cdot)\|_{H^{\sigma}(\mathbb{S})}^{2}
\lesssim (\|u_{\omega, n}\|_{H^{s}(\mathbb{S})}+\|u^{\omega, n}\|_{H^{s}(\mathbb{S})})\|v\|_{H^{\sigma}(\mathbb{S})}^{2}
+\|F\|_{H^{\sigma}(\mathbb{S})}\|v\|_{H^{\sigma}(\mathbb{S})}.
\end{array}
$$
By (4.1), we have
$$
\begin{array}{rl}
\|u_{\omega, n}(t, \cdot)\|_{H^{s}(\mathbb{S})}+\|u^{\omega, n}(t, \cdot)\|_{H^{s}(\mathbb{S})}
\leq e^{cT_{0}}\|u_{\omega, n}(0, \cdot)\|_{H^{s}(\mathbb{S})}+\|u^{\omega, n}(t, \cdot)\|_{H^{s}(\mathbb{S})}\lesssim 1, \quad t\in[0, T_{0}].
\end{array}
$$
According to Lemma 4.5, we obtain
$$
\begin{array}{rl}
\frac{1}{2}\frac{d}{dt}\|v(t, \cdot)\|_{H^{\sigma}(\mathbb{S})}^{2}
\lesssim \|v\|_{H^{\sigma}(\mathbb{S})}^{2}
+n^{-r_{s}}\|v\|_{H^{\sigma}(\mathbb{S})}.
\end{array}
$$
That is,
$$
\begin{array}{rl}
\frac{d}{dt}\|v(t, \cdot)\|_{H^{\sigma}(\mathbb{S})}
\lesssim \|v\|_{H^{\sigma}(\mathbb{S})}
+n^{-r_{s}}.
\end{array}
$$
Since $v(0,x)=0$, the Gronwall's inequality implies the desired result. \hfill $\Box$

\subsection{Non-uniform dependence}

The following theorem is our main result in this section.\\

\noindent\textbf{Theorem 4.2.} If $s\geq 4$, then for any fixed $T_{0}>0$, the solution map $u_{0}\rightarrow u(t)$ of Eq. (2.4) is not uniformly continuous from any bounded subset of $H^{s}(\mathbb{S})$ into
$C([0, T_{0}]; H^{s}(\mathbb{S}))$. More precisely, there exist two
sequences of $u_{n}(t)$ and $v_{n}(t)$ in $C([0, T_{0}]; H^{s}(\mathbb{S}))$ such that
$$
\begin{array}{rl}
\|u_{n}(t)\|_{H^{s}(\mathbb{S})}+\|v_{n}(t)\|_{H^{s}(\mathbb{S})}\lesssim 1,\\[5pt]
\lim_{n\rightarrow \infty}\|u_{n}(0)-v_{n}(0)\|_{H^{s}(\mathbb{S})}=0,
\end{array}
$$
and
$$
\begin{array}{rl}
\liminf_{n\rightarrow \infty}\|u_{n}(t)-v_{n}(t)\|_{H^{s}(\mathbb{S})}
\gtrsim |\sin t|, \quad t\in[0, T_{0}].
\end{array}
$$

\noindent\textbf{Proof.} Let $u_{1, n}(t, x)$ and $u_{-1, n}(t, x)$ be the unique solutions to Eq. (2.4) with the initial data $u_{1, n}(0, x)$ and $u_{-1, n}(0, x)$, respectively. Using Lemma 4.1, we have
$$
\begin{array}{rl}
\|u_{1, n}(t)\|_{H^{s}(\mathbb{S})}+\|u_{-1, n}(t)\|_{H^{s}(\mathbb{S})}
\lesssim \|u_{1, n}(0)\|_{H^{s}(\mathbb{S})}+\|u_{-1, n}(0)\|_{H^{s}(\mathbb{S})}
\lesssim 1.
\end{array}
$$
Moreover,
$$
\begin{array}{rl}
\lim_{n\rightarrow \infty}\|u_{1, n}(0)-u_{-1, n}(0)\|_{H^{s}(\mathbb{S})}
=\lim_{n\rightarrow \infty}\|2n^{-1}\|_{H^{s}(\mathbb{S})}
=0.
\end{array}
$$
Since $2s-\sigma\geq s\geq 4$, applying Lemma 4.6 and the interpolation inequality
$$
\|f\|_{H^{s}(\mathbb{S})}
\leq \|f\|_{H^{\sigma}(\mathbb{S})}^{\frac{1}{2}}
\|f\|_{H^{2s-\sigma}(\mathbb{S})}^{\frac{1}{2}},
$$
we have
$$
\begin{array}{rl}
&\|u^{\pm 1, n}(t)-u_{\pm1, n}(t)\|_{H^{s}(\mathbb{S})}\\[5pt]
&\lesssim \|u^{\pm 1, n}(t)-u_{\pm1, n}(t)\|_{H^{\sigma}(\mathbb{S})}^{\frac{1}{2}}
\|u^{\pm 1, n}(t)-u_{\pm1, n}(t)\|_{H^{2s-\sigma}(\mathbb{S})}^{\frac{1}{2}}\\[5pt]
&\lesssim n^{-\frac{1}{2}r_{s}}[\|u^{\pm 1, n}(t)\|_{H^{2s-\sigma}(\mathbb{S})}
+\|u_{\pm1, n}(t)\|_{H^{2s-\sigma}(\mathbb{S})}]^{\frac{1}{2}}\\[5pt]
&\lesssim n^{-\frac{1}{2}r_{s}}
[\|\pm n^{-1}+n^{-s}\cos(nx\mp t)\|_{H^{2s-\sigma}(\mathbb{S})}
+\|u_{\pm1, n}(0)\|_{H^{2s-\sigma}(\mathbb{S})}]^{\frac{1}{2}}\\[5pt]
&=n^{-\frac{1}{2}r_{s}}
[\|\pm n^{-1}+n^{-s}\cos(nx\mp t)\|_{H^{2s-\sigma}(\mathbb{S})}
+\|\pm n^{-1}+n^{-s}\cos(nx)\|_{H^{2s-\sigma}(\mathbb{S})}]^{\frac{1}{2}}\\[5pt]
&\lesssim n^{-\frac{1}{2}r_{s}}(n^{-1}+n^{s-\sigma})^{\frac{1}{2}}.
\end{array}
\eqno(4.4)
$$
By Lemma 4.2,
$$
\begin{array}{rl}
&\liminf_{n\rightarrow \infty}\|u^{1, n}(t)-u^{-1, n}(t)\|_{H^{s}(\mathbb{S})}\\[5pt]
&=\liminf_{n\rightarrow \infty}\|2n^{-1}+n^{-s}[\cos(nx+t)-\cos(nx-t)]\|_{H^{s}(\mathbb{S})}\\[5pt]
&=\liminf_{n\rightarrow \infty}\|2n^{-1}+2n^{-s}\sin(nx)\sin~t\|_{H^{s}(\mathbb{S})}\\[5pt]
&\geq \liminf_{n\rightarrow \infty}
(\|2n^{-s}\sin(nx)\sin~t\|_{H^{s}(\mathbb{S})}-\|2n^{-1}\|_{H^{s}(\mathbb{S})})\\[5pt]
&\gtrsim |\sin~t|.
\end{array}
\eqno(4.5)
$$
Therefore, by (4.4) and (4.5), we know
$$
\begin{array}{rl}
&\liminf_{n\rightarrow \infty}\|u_{1, n}(t)-u_{-1, n}(t)\|_{H^{s}(\mathbb{S})}\\[5pt]
&\geq\liminf_{n\rightarrow \infty}
(\|u^{1, n}(t)-u^{-1, n}(t)\|_{H^{s}(\mathbb{S})}
-\|u^{1, n}(t)-u_{1, n}(t)\|_{H^{s}(\mathbb{S})}
-\|u^{-1, n}(t)-u_{-1, n}(t)\|_{H^{s}(\mathbb{S})})\\[5pt]
&\geq\liminf_{n\rightarrow \infty}
\|u^{1, n}(t)-u^{-1, n}(t)\|_{H^{s}(\mathbb{S})}\\[5pt]
&\quad-\lim_{n\rightarrow \infty}(\|u^{1, n}(t)-u_{1, n}(t)\|_{H^{s}(\mathbb{S})}
+\|u^{-1, n}(t)-u_{-1, n}(t)\|_{H^{s}(\mathbb{S})})\\[5pt]
&\gtrsim |\sin~t|,
\end{array}
$$
which completes the proof.   \hfill $\Box$ \\

\noindent\textbf{Remark 4.1.} If we consider the solution in a small time interval, we can extend the condition $s\geq 4$ in Theorem 4.1 to $s>\frac{7}{2}$. In fact, the restriction condition $s\geq 4$ is assumed in Lemma 4.1 to get the estimate (4.1). For $s>\frac{7}{2}$, by (3.6), we know
$$
\begin{array}{l}
\frac{1}{2}\frac{d}{dt}\|u\|_{H^{s}(\mathbb{S})}^{2}
\leq c_{s}(\|u\|_{H^{s}(\mathbb{S})}^{2}+\|u\|_{H^{s}(\mathbb{S})}^{3}),
\end{array}
$$
which implies
$$
\begin{array}{l}
\|u(t)\|_{H^{s}(\mathbb{S})}
\leq 2e^{c_{s}T_{0}}\|u_{0}\|_{H^{s}(\mathbb{S})}, \quad
\forall t\in [0, T_{0}] ~\mbox{with}~T_{0}:=\frac{1}{2c}\ln(1+\frac{1}{\|u_{0}\|_{H^{s}(\mathbb{S})}}).
\end{array}
$$

\section{Global existence of weak solution}

In this section, we establish the existence of global weak solution in $H^{2}(\mathbb{S})$.
Firstly, the Cauchy problem (2.4) can be rewritten as follows
$$
\begin{array}{l}
\left\{\begin{array}{l}
\partial_{t}u+u\partial_{x}u
 +\partial_{x}P=0,\quad t>0, ~x\in \mathbb{R},\\[5pt]
A_{\mu}P=2\mu(u)u+\frac{1}{2}(\partial_{x}u)^{2}-\frac{1}{2}(\partial_{x}^{2}u)^{2}
 -3\partial_{x}(\partial_{x}u\partial_{x}^{2}u),\\[5pt]
u(t,x+1)=u(t,x), \quad t\geq0, ~x\in \mathbb{R},\\[5pt]
u(0,x)=u_{0}(x), \quad x\in \mathbb{R}.
\end{array}
\right.
\end{array}
\eqno(5.1)
$$
Now we introduce the definition of a weak solution to the Cauchy problem (5.1).\\

\noindent\textbf{Definition 5.1.} We call
$u: \mathbb{R}_{+}\times \mathbb{S}\rightarrow \mathbb{R}$ an admissible global weak solution of the Cauchy problem (5.1) if\\
$(i)$ $u(t,x)\in C(\mathbb{R}_{+}; C^{1}(\mathbb{S}))\cap L^{\infty}(\mathbb{R}_{+}; H^{2}(\mathbb{S}))$
and
$$
\begin{array}{l}
\|\partial_{x}u(t,\cdot)\|_{H^{1}(\mathbb{S})}
\leq\|\partial_{x}u_{0}\|_{H^{1}(\mathbb{S})} \quad \mbox{for each}~ t>0.
\end{array}
\eqno(5.2)
$$
$(ii)$ $u(t,x)$ satisfies Eq. (5.1) in the sense of distributions and takes on the initial data pointwise.\\

The main result of this section is as follows.\\

\noindent\textbf{Theorem 5.1.} Let $p>2$. For any $u_{0}\in H^{2}(\mathbb{S})$ satisfying $\partial_{x}^{2}u_{0}\in L^{p}(\mathbb{S})$,
the Cauchy problem (5.1) has an admissible global weak solution in the sense of Definition 5.1.

\subsection{\textbf{Viscous approximate solutions}}

In this subsection, we construct the approximation solution sequence
$u_{\varepsilon}=u_{\varepsilon}(t,x)$. Hence, we consider the viscous problem of
Eq. (5.1) as follows
$$
\left\{\begin{array}{rl}
&\partial_{t}u_{\varepsilon}+u_{\varepsilon}\partial_{x}u_{\varepsilon}
+\partial_{x}P_{\varepsilon}=\varepsilon\partial_{x}^{2}u_{\varepsilon},
 \quad t>0,~x\in \mathbb{R},\\[3pt]
&A_{\mu}P_{\varepsilon}=2\mu(u_{\varepsilon})u_{\varepsilon}
 +\frac{1}{2}(\partial_{x}u_{\varepsilon})^{2}
 -\frac{1}{2}(\partial_{x}^{2}u_{\varepsilon})^{2}
 -3\partial_{x}(\partial_{x}u_{\varepsilon}\partial_{x}^{2}u_{\varepsilon}),
 \quad t\geq0,~x\in \mathbb{R},\\[3pt]
&u_{\varepsilon}(t,x+1)=u_{\varepsilon}(t,x), \quad t\geq0,~x\in \mathbb{R},\\[3pt]
&u_{\varepsilon}(0,x)=u_{\varepsilon,0}(x), \quad x\in \mathbb{R},
\end{array}
\right.
\eqno(5.3)
$$
where $u_{\varepsilon,0}(x)=(j_{\varepsilon}*u_{0})(x)$ and
$J_{\varepsilon}, j_{\varepsilon}$ are defined in (3.1).
By Lemma 3.3, we have
$$
\begin{array}{rl}
\|u_{\varepsilon,0}\|_{L^{2}(\mathbb{S})}\leq
\|u_{0}\|_{L^{2}(\mathbb{S})},~
\|\partial_{x}u_{\varepsilon,0}\|_{L^{2}(\mathbb{S})}\leq \|\partial_{x}u_{0}\|_{L^{2}(\mathbb{S})},\\[5pt]
\|\partial_{x}^{2}u_{\varepsilon,0}\|_{L^{2}(\mathbb{S})}\leq \|\partial_{x}^{2}u_{0}\|_{L^{2}(\mathbb{S})},~
\|\partial_{x}^{2}u_{\varepsilon,0}\|_{L^{p}(\mathbb{S})}\leq \|\partial_{x}^{2}u_{0}\|_{L^{p}(\mathbb{S})}
\end{array}
\eqno(5.4)
$$
and $u_{\varepsilon,0}\rightarrow u_{0} ~ \mbox{in}~H^{2}(\mathbb{S})$ as $\varepsilon\rightarrow 0$.\\

\noindent\textbf{Lemma 5.1.}
Let $\varepsilon>0$ and $u_{\varepsilon,0}\in H^{s}(\mathbb{S})$, $s\geq 5$. Then
there exists a unique
$u_{\varepsilon}\in C(\mathbb{R}_{+}; H^{s}(\mathbb{S}))$
to Eq. (5.3). Moreover, for each $t\geq 0$ and $\varepsilon>0$, it holds that
$$
\begin{array}{rl}
&\int_{\mathbb{S}}[(\partial_{x}u_{\varepsilon})^{2}
+(\partial_{x}^{2} u_{\varepsilon})^{2}](t,x)dx
+2\varepsilon\int_{0}^{t}\int_{\mathbb{S}}
[(\partial_{x}^{2} u_{\varepsilon})^{2}
+(\partial_{x}^{3} u_{\varepsilon})^{2}](s,x)dxds\\[3pt]
&\quad=\int_{\mathbb{S}}[(\partial_{x}u_{\varepsilon,0})^{2}
+(\partial_{x}^{2} u_{\varepsilon,0})^{2}]dx,
\end{array}
\eqno(5.5)
$$
and for each $\varepsilon>0$,
$$
\begin{array}{rl}
\|u_{\varepsilon}\|_{L^{\infty}(\mathbb{R}_{+}\times \mathbb{S})},~
\|\partial_{x}u_{\varepsilon}\|_{L^{\infty}(\mathbb{R}_{+}\times \mathbb{S})}
\leq \|u_{0}\|_{H^{2}(\mathbb{S})}.
\end{array}
$$

\noindent\textbf{Proof.}
First, following the standard argument for a nonlinear parabolic equation, one can obtain
the local well-posedness result that, for $u_{\varepsilon,0}\in H^{s}(\mathbb{S})$, $s\geq4$,
there exists a positive constant $T_{0}$ such that Eq. (5.3) has a unique solution
$u_{\varepsilon}=u_{\varepsilon}(t,x)\in C([0,T_{0}]; H^{s}(\mathbb{S}))\cap L^{2}([0,T_{0}]; H^{s+1}(\mathbb{S}))$. We denote the life span of the solution $u_{\varepsilon}(t,x)$ by $T$.
Note that
$(A_{\mu}f, g)_{L^{2}(\mathbb{S})}=(f, A_{\mu}g)_{L^{2}(\mathbb{S})}$.
Multiplying Eq. (5.3) by $A_{\mu}u_{\varepsilon}$ and integrating over $\mathbb{S}$, we obtain
$$
\frac{1}{2}\frac{d}{dt}\int_{\mathbb{S}}[(\partial_{x}u_{\varepsilon})^{2}
+(\partial_{x}^{2} u_{\varepsilon})^{2}](t,x)dx
=-\varepsilon\int_{\mathbb{S}}[(\partial_{x}^{2} u_{\varepsilon})^{2}
+(\partial_{x}^{3} u_{\varepsilon})^{2}](t,x)dx.
$$
Then $(5.5)$ holds for all $0\leq t<T$.

Similar to the proof of Theorem 3.1, we can show that if
$\|\partial_{x}^{3}u_{\varepsilon}(t,\cdot)\|_{L^{\infty}(\mathbb{S})}
<\infty$, then the $H^{s}(\mathbb{S})$-norm of $u_{\varepsilon}(t,\cdot)$ does not blow up on $[0,T)$. Next, we prove $T=\infty$.
By Corollary 2.1, (5.4) and (5.5), we obtain that
$$
\begin{array}{rl}
\max_{x\in\mathbb{S}}(\partial_{x}u_{\varepsilon})^{2}(t,x)
&\leq \frac{1}{12}\int_{\mathbb{S}}(\partial_{x}^{2}u_{\varepsilon})^{2}(t,x)dx
\leq\frac{1}{12}\int_{\mathbb{S}}[(\partial_{x} u_{\varepsilon,0})^{2}
+(\partial_{x}^{2} u_{\varepsilon,0})^{2}]dx\\[5pt]
&\leq\frac{1}{12}(\|\partial_{x}u_{0}\|^{2}_{L^{2}(\mathbb{S})}
+\|\partial_{x}^{2}u_{0}\|^{2}_{L^{2}(\mathbb{S})}).
\end{array}
$$
This in turn implies that
$$
\begin{array}{l}
\|\partial_{x}u_{\varepsilon}\|_{L^{\infty}(\mathbb{S})}
\leq \frac{\sqrt{3}}{6}\mu_{1},
\end{array}
$$
where $\mu_{1}$ is defined in (2.5).
Note that
$
\int_{\mathbb{S}}(u_{\varepsilon}(t,x)-\mu_{\varepsilon,0})=0.
$
By Corollary 2.1,
$$
\begin{array}{rl}
\max_{x\in\mathbb{S}}(u_{\varepsilon}(t,x)-\mu_{\varepsilon,0})^{2}
\leq \frac{1}{12}\int_{\mathbb{S}}(\partial_{x}u_{\varepsilon})^{2}(t,x)dx
\leq \frac{1}{12}\|\partial_{x}u_{\varepsilon}(t,\cdot)\|_{L^{\infty}(\mathbb{S})}^{2}.
\end{array}
$$
Hence, we get
$$
\|u_{\varepsilon}\|_{L^{\infty}(\mathbb{S})}
\leq \frac{1}{12}\mu_{1}
+|\mu_{\varepsilon,0}|
\leq \frac{1}{12}\mu_{1}
+\|u_{0}\|_{L^{2}(\mathbb{S})}.
$$

Due to Corollary 2.1, we only need to derive an a priori estimate on
$\|\partial_{x}^{4}u_{\varepsilon}\|_{L^{2}(\mathbb{S})}$. Applying the operator $A_{\mu}$ to Eq. (5.3), then multiplying both sides by $A_{\mu}u_{\varepsilon}$ and integrating over $\mathbb{S}$
with respect to $x$, we get
$$
\begin{array}{rl}
&\frac{1}{2}\frac{d}{dt}\int_{\mathbb{S}}[
(\partial_{x}^{2} u_{\varepsilon})^{2}
+2(\partial_{x}^{3} u_{\varepsilon})^{2}
+(\partial_{x}^{4} u_{\varepsilon})^{2}](t,x)dx
+\varepsilon\int_{\mathbb{S}}[
(\partial_{x}^{3} u_{\varepsilon})^{2}
+2(\partial_{x}^{4} u_{\varepsilon})^{2}
+(\partial_{x}^{5} u_{\varepsilon})^{2}](t,x)dx\\[5pt]
&=
-\int_{\mathbb{S}}u_{\varepsilon}\partial_{x}u_{\varepsilon}\partial_{x}^{4}u_{\varepsilon}dx
+\int_{\mathbb{S}}\partial_{x}u_{\varepsilon}(\partial_{x}^{2}u_{\varepsilon})^{2}dx
-\int_{\mathbb{S}}\partial_{x}u_{\varepsilon}(\partial_{x}^{4}u_{\varepsilon})^{2}dx\\[5pt]
&\quad
+2\int_{\mathbb{S}}\partial_{x}u_{\varepsilon}\partial_{x}^{2}u_{\varepsilon}\partial_{x}^{4}u_{\varepsilon}dx
+2\int_{\mathbb{S}}u_{\varepsilon}\partial_{x}^{3}u_{\varepsilon}\partial_{x}^{4}u_{\varepsilon}dx
+\int_{\mathbb{S}}u_{\varepsilon}\partial_{x}^{4}u_{\varepsilon}\partial_{x}^{5}u_{\varepsilon}dx\\[5pt]
&\leq\|u_{\varepsilon}\|_{L^{\infty}(\mathbb{S})}\|\partial_{x}u_{\varepsilon}\|_{L^{\infty}(\mathbb{S})}
\|\partial_{x}^{4} u_{\varepsilon}\|_{L^{2}(\mathbb{S})}
+\|\partial_{x}u_{\varepsilon}\|_{L^{\infty}(\mathbb{S})}
\|\partial_{x}^{2} u_{\varepsilon}\|^{2}_{L^{2}(\mathbb{S})}
+\|\partial_{x}u_{\varepsilon}\|_{L^{\infty}(\mathbb{S})}
\|\partial_{x}^{4} u_{\varepsilon}\|^{2}_{L^{2}(\mathbb{S})}\\[5pt]
&\quad
+2\|\partial_{x}u_{\varepsilon}\|_{L^{\infty}(\mathbb{S})}
\|\partial_{x}^{2}u_{\varepsilon}\|_{L^{2}(\mathbb{S})}
\|\partial_{x}^{4} u_{\varepsilon}\|_{L^{2}(\mathbb{S})}
+2\|u_{\varepsilon}\|_{L^{\infty}(\mathbb{S})}
\|\partial_{x}^{3}u_{\varepsilon}\|_{L^{2}(\mathbb{S})}
\|\partial_{x}^{4} u_{\varepsilon}\|_{L^{2}(\mathbb{S})}\\[5pt]
&\quad
+\varepsilon\|\partial_{x}^{5} u_{\varepsilon}\|_{L^{2}(\mathbb{S})}^{2}
+\frac{1}{4\varepsilon}\|u_{\varepsilon}\|_{L^{\infty}(\mathbb{S})}^{2}
\|\partial_{x}^{4} u_{\varepsilon}\|_{L^{2}(\mathbb{S})}^{2},
\end{array}
$$
which implies that $\|\partial_{x}^{4} u_{\varepsilon}\|_{L^{2}(\mathbb{S})}$ is bounded for $t\in [0, T)$.
Since $\|\partial_{x}^{3}u_{\varepsilon}\|_{L^{\infty}(\mathbb{S})}
\leq \frac{\sqrt{3}}{6}\|\partial_{x}^{4} u_{\varepsilon}\|_{L^{2}(\mathbb{S})}$ by Corollary 2.1, we have $T=\infty$, which completes the proof of the lemma. \hfill $\Box$

\subsection{\textbf{Precompactness}}

In this subsection, we are ready to obtain the necessary compactness of the viscous approximation solutions $u_{\varepsilon}(t,x)$.

For convenience, we denote $P_{\varepsilon}=P_{1,\varepsilon}+P_{2,\varepsilon}$, where $P_{1,\varepsilon}, P_{2,\varepsilon}$ are defined by
$$
\begin{array}{rl}
&P_{1,\varepsilon}
=A_{\mu}^{-1}
[2\mu(u_{\varepsilon})u_{\varepsilon}+\frac{1}{2}(\partial_{x}u_{\varepsilon})^{2}
-\frac{1}{2}(\partial_{x}^{2}u_{\varepsilon})^{2}],\\[5pt]
&P_{2,\varepsilon}
=
-3\partial_{x}A_{\mu}^{-1}(\partial_{x}u_{\varepsilon}\partial_{x}^{2}u_{\varepsilon}).
\end{array}
$$

\noindent\textbf{Lemma 5.2.} Assume $u_{0}\in H^{2}(\mathbb{S})$. For each $t\geq 0$ and $\varepsilon>0$, the following inequalities hold
$$
\begin{array}{rl}
&\|P_{1,\varepsilon}(t, \cdot)\|_{W^{4,1}(\mathbb{S})},~
\|P_{1,\varepsilon}(t, \cdot)\|_{W^{3,\infty}(\mathbb{S})}
\leq C_{0}\|u_{0}\|_{H^{2}(\mathbb{S})}^{2},\\[5pt]
&\|P_{2,\varepsilon}(t, \cdot)\|_{W^{2,1}(\mathbb{S})},~
\|P_{2,\varepsilon}(t, \cdot)\|_{W^{2,\infty}(\mathbb{S})}
\leq C_{0}\|u_{0}\|_{H^{2}(\mathbb{S})}^{2},\\[5pt]
&\|P_{\varepsilon}(t, \cdot)\|_{W^{2,1}(\mathbb{S})},~
\|P_{\varepsilon}(t, \cdot)\|_{W^{2,\infty}(\mathbb{S})}
\leq C_{0}\|u_{0}\|_{H^{2}(\mathbb{S})}^{2},\\[5pt]
&\|\partial_{x}^{3}P_{\varepsilon}(t, \cdot)\|_{L^{1}(\mathbb{S})}
\leq C_{0}\|u_{0}\|_{H^{2}(\mathbb{S})}^{2}.
\end{array}
$$
Here and in what follows, we use $C_{0}$ to denote a generic positive constant, independent of $\varepsilon$, which may change from line to line.\\

\noindent\textbf{Proof.} For $\sigma=1$ or $\infty$, by Young's inequality, we have
$$
\begin{array}{rl}
\|\partial_{x}^{i}P_{1,\varepsilon}(t, \cdot)\|_{L^{\sigma}(\mathbb{S})}
&=\|\partial_{x}^{i}A_{\mu}^{-1}
[2\mu(u_{\varepsilon})u_{\varepsilon}+\frac{1}{2}(\partial_{x}u_{\varepsilon})^{2}
-\frac{1}{2}(\partial_{x}^{2}u_{\varepsilon})^{2}]\|_{L^{\sigma}(\mathbb{S})}\\[5pt]
&=\|\partial_{x}^{i}g*[2\mu(u_{\varepsilon})u_{\varepsilon}+\frac{1}{2}(\partial_{x}u_{\varepsilon})^{2}
-\frac{1}{2}(\partial_{x}^{2}u_{\varepsilon})^{2}]\|_{L^{\sigma}(\mathbb{S})}\\[5pt]
&\leq \|\partial_{x}^{i}g\|_{L^{\sigma}(\mathbb{S})}
\|2\mu(u_{\varepsilon})u_{\varepsilon}+\frac{1}{2}(\partial_{x}u_{\varepsilon})^{2}
-\frac{1}{2}(\partial_{x}^{2}u_{\varepsilon})^{2}\|_{L^{1}(\mathbb{S})}\\[5pt]
&\leq C_{0}\|u_{0}\|_{H^{2}(\mathbb{S})}^{2}.
\end{array}
$$
and
$$
\begin{array}{rl}
\|\partial_{x}^{j}P_{2,\varepsilon}(t, \cdot)\|_{L^{\sigma}(\mathbb{S})}
&=\|-3\partial_{x}^{j+1}A_{\mu}^{-1}(\partial_{x}u_{\varepsilon}\partial_{x}^{2}u_{\varepsilon})\|_{L^{\sigma}(\mathbb{S})}\\[5pt]
&=\|-3\partial_{x}^{j+1}g*(\partial_{x}u_{\varepsilon}\partial_{x}^{2}u_{\varepsilon})\|_{L^{\sigma}(\mathbb{S})}\\[5pt]
&\leq 3\|\partial_{x}^{j+1}g\|_{L^{\sigma}(\mathbb{S})}
\|\partial_{x}u_{\varepsilon}\partial_{x}^{2}u_{\varepsilon}\|_{L^{1}(\mathbb{S})}\\[5pt]
&\leq C_{0}\|u_{0}\|_{H^{2}(\mathbb{S})}^{2},
\end{array}
$$
where $i=0,1,2,3$ and $j=0,1,2$.
The estimate of $\|\partial_{x}^{4}P_{1,\varepsilon}\|_{L^{1}(\mathbb{S})}$ follows from the above estimates and the fact
$$
\begin{array}{rl}
\partial_{x}^{4}P_{1,\varepsilon}
=-\mu(P_{1,\varepsilon})+\partial_{x}^{2}P_{1,\varepsilon}
+2\mu(u_{\varepsilon})u_{\varepsilon}+\frac{1}{2}(\partial_{x}u_{\varepsilon})^{2}
-\frac{1}{2}(\partial_{x}^{2}u_{\varepsilon})^{2}.
\end{array}
$$
Since $P_{\varepsilon}=P_{1,\varepsilon}+P_{2,\varepsilon}$, we can directly deduce the estimate of $\|P_{\varepsilon}(t, \cdot)\|_{W^{2,\sigma}(\mathbb{S})}$.

Moreover,
$$
\begin{array}{rl}
\partial_{x}^{3}P_{\varepsilon}
&=\partial_{x}^{3}P_{1,\varepsilon}+\partial_{x}^{3}P_{2,\varepsilon}
=\partial_{x}^{3}P_{1,\varepsilon}
-3\partial_{x}^{4}A_{\mu}^{-1}(\partial_{x}u_{\varepsilon}\partial_{x}^{2}u_{\varepsilon})\\[5pt]
&=\partial_{x}^{3}P_{1,\varepsilon}
-3\partial_{x}u_{\varepsilon}\partial_{x}^{2}u_{\varepsilon}
+3\mu(A_{\mu}^{-1}(\partial_{x}u_{\varepsilon}\partial_{x}^{2}u_{\varepsilon}))
-3\partial_{x}^{2}A_{\mu}^{-1}(\partial_{x}u_{\varepsilon}\partial_{x}^{2}u_{\varepsilon}),
\end{array}
$$
then by Young's inequality, we have
$$
\begin{array}{rl}
\|\partial_{x}^{3}P_{\varepsilon}\|_{L^{1}(\mathbb{S})}
&\leq \|\partial_{x}^{3}P_{1,\varepsilon}\|_{L^{1}(\mathbb{S})}
+3\|\partial_{x}u_{\varepsilon}\partial_{x}^{2}u_{\varepsilon}\|_{L^{1}(\mathbb{S})}\\[5pt]
&\quad
+3\|A_{\mu}^{-1}(\partial_{x}u_{\varepsilon}\partial_{x}^{2}u_{\varepsilon})\|_{L^{1}(\mathbb{S})}
+3\|\partial_{x}^{2}A_{\mu}^{-1}(\partial_{x}u_{\varepsilon}\partial_{x}^{2}u_{\varepsilon})\|_{L^{1}(\mathbb{S})}\\[5pt]
&\leq \|\partial_{x}^{3}P_{1,\varepsilon}\|_{L^{1}(\mathbb{S})}
+3\|\partial_{x}u_{\varepsilon}\|_{L^{\infty}(\mathbb{S})}
\|\partial_{x}^{2}u_{\varepsilon}\|_{L^{2}(\mathbb{S})}\\[5pt]
&\quad
+3\|g\|_{L^{1}(\mathbb{S})}\|\partial_{x}u_{\varepsilon}\partial_{x}^{2}u_{\varepsilon}\|_{L^{1}(\mathbb{S})}
+3\|\partial_{x}^{2}g\|_{L^{1}(\mathbb{S})}\|\partial_{x}u_{\varepsilon}\partial_{x}^{2}u_{\varepsilon}\|_{L^{1}(\mathbb{S})}\\[5pt]
&\leq C_{0}\|u_{0}\|_{H^{2}(\mathbb{S})}^{2},
\end{array}
$$
which completes the proof. \hfill $\Box$\\

Next we turn to estimates of time derivatives. \\

\noindent\textbf{Lemma 5.3.} Assume $u_{0}\in H^{2}(\mathbb{S})$. For each $T, t>0$ and $0<\varepsilon<1$,
the following inequalities hold
$$
\begin{array}{rl}
&\|\partial_{t}u_{\varepsilon}(t, \cdot)\|_{L^{2}(\mathbb{S})}
\leq C_{0}\|u_{0}\|_{H^{2}(\mathbb{S})}^{2}
+\|u_{0}\|_{H^{2}(\mathbb{S})}, \\[5pt]
&\|\partial_{t}\partial_{x}u_{\varepsilon}\|_{L^{2}([0, T]\times\mathbb{S})}
\leq C_{0}\sqrt{T}\|u_{0}\|_{H^{2}(\mathbb{S})}^{2}
+\frac{\sqrt{2}}{2} \|u_{0}\|_{H^{2}(\mathbb{S})}, \\[5pt]
&\|\partial_{t}\partial_{x}^{3}P_{1,\varepsilon}\|_{L^{1}([0, T]\times\mathbb{S})}
\leq C_{0}(T+1)(\|u_{0}\|_{H^{2}(\mathbb{S})}^{3}+\|u_{0}\|_{H^{2}(\mathbb{S})}^{2}).
\end{array}
$$

\noindent\textbf{Proof.} By the first equation of (5.3) and Lemmas 5.1-5.2, we have
$$
\begin{array}{rl}
\|\partial_{t}u_{\varepsilon}(t, \cdot)\|_{L^{2}(\mathbb{S})}
&\leq \|u_{\varepsilon}\partial_{x}u_{\varepsilon}\|_{L^{2}(\mathbb{S})}
+\|\partial_{x}P_{\varepsilon}\|_{L^{2}(\mathbb{S})}
+\varepsilon\|\partial_{x}^{2}u_{\varepsilon}\|_{L^{2}(\mathbb{S})}\\[5pt]
&\leq C_{0}\|u_{0}\|_{H^{2}(\mathbb{S})}^{2}
+\varepsilon \|u_{0}\|_{H^{2}(\mathbb{S})}
\leq C_{0}\|u_{0}\|_{H^{2}(\mathbb{S})}^{2}
+\|u_{0}\|_{H^{2}(\mathbb{S})}.
\end{array}
\eqno(5.6)
$$
Differentiating the first equation of (5.3) with respect to $x$, one obtains
$$
\begin{array}{rl}
\partial_{t}\partial_{x}u_{\varepsilon}
+u_{\varepsilon}\partial_{x}^{2}u_{\varepsilon}
+(\partial_{x}u_{\varepsilon})^{2}
+\partial_{x}^{2}P_{\varepsilon}
=\varepsilon\partial_{x}^{3}u_{\varepsilon}.
\end{array}
$$
Thus,
$$
\begin{array}{rl}
\|\partial_{t}\partial_{x}u_{\varepsilon}\|_{L^{2}([0, T]\times\mathbb{S})}
&\leq \|u_{\varepsilon}\partial_{x}^{2}u_{\varepsilon}\|_{L^{2}([0, T]\times\mathbb{S})}
+\|\partial_{x}u_{\varepsilon}\|_{L^{4}([0, T]\times\mathbb{S})}^{2}\\[5pt]
&\quad +\|\partial_{x}^{2}P_{\varepsilon}\|_{L^{2}([0, T]\times\mathbb{S})}
+\varepsilon\|\partial_{x}^{3}u_{\varepsilon}\|_{L^{2}([0, T]\times\mathbb{S})}\\[5pt]
&\leq \|u_{\varepsilon}\|_{L^{\infty}([0, T]\times\mathbb{S})}\|\partial_{x}^{2}u_{\varepsilon}\|_{L^{2}([0, T]\times\mathbb{S})}
+\sqrt{T}\|\partial_{x}u_{\varepsilon}\|_{L^{\infty}([0, T]\times\mathbb{S})}^{2}\\[5pt]
&\quad +\sqrt{T}\|\partial_{x}^{2}P_{\varepsilon}\|_{L^{\infty}([0, T]\times\mathbb{S})}
+\varepsilon\|\partial_{x}^{3}u_{\varepsilon}\|_{L^{2}([0, T]\times\mathbb{S})}\\[5pt]
&\leq C_{0}\sqrt{T}\|u_{0}\|_{H^{2}(\mathbb{S})}^{2}
+\frac{\sqrt{2}}{2} \|u_{0}\|_{H^{2}(\mathbb{S})}.
\end{array}
\eqno(5.7)
$$
Moreover,
differentiating the first equation of (5.3) with respect to $x$ two times, we have
$$
\begin{array}{rl}
\partial_{t}\partial_{x}^{2}u_{\varepsilon}
+u_{\varepsilon}\partial_{x}^{3}u_{\varepsilon}
+3\partial_{x}u_{\varepsilon}\partial_{x}^{2}u_{\varepsilon}
+\partial_{x}^{3}P_{\varepsilon}
=\varepsilon\partial_{x}^{4}u_{\varepsilon},
\end{array}
$$
and then
$$
\begin{array}{rl}
-\partial_{x}^{2}u_{\varepsilon}\partial_{t}\partial_{x}^{2}u_{\varepsilon}
&=\partial_{x}^{2}u_{\varepsilon}(u_{\varepsilon}\partial_{x}^{3}u_{\varepsilon}
+3\partial_{x}u_{\varepsilon}\partial_{x}^{2}u_{\varepsilon}
+\partial_{x}^{3}P_{\varepsilon}
-\varepsilon\partial_{x}^{4}u_{\varepsilon})\\[5pt]
&=\frac{1}{2}\partial_{x}(u_{\varepsilon}(\partial_{x}^{2}u_{\varepsilon})^{2})
-\frac{1}{2}\partial_{x}u_{\varepsilon}(\partial_{x}^{2}u_{\varepsilon})^{2}
+\partial_{x}^{2}u_{\varepsilon}\partial_{x}^{3}P_{1,\varepsilon}
+3\partial_{x}^{2}u_{\varepsilon}\mu(A_{\mu}^{-1}(\partial_{x}u_{\varepsilon}\partial_{x}^{2}u_{\varepsilon}))\\[5pt]
&\quad
-3\partial_{x}^{2}u_{\varepsilon}\partial_{x}^{2}A_{\mu}^{-1}(\partial_{x}u_{\varepsilon}\partial_{x}^{2}u_{\varepsilon})
-\varepsilon\partial_{x}(\partial_{x}^{2}u_{\varepsilon}\partial_{x}^{3}u_{\varepsilon})
+\varepsilon(\partial_{x}^{3}u_{\varepsilon})^{2}.
\end{array}
$$
By the definition of $P_{1,\varepsilon}$, we know
$$
\begin{array}{rl}
\partial_{t}\partial_{x}^{3}P_{1,\varepsilon}
&=\partial_{x}^{3}A_{\mu}^{-1}
[2\mu(u_{\varepsilon})\partial_{t}u_{\varepsilon}
+\partial_{x}u_{\varepsilon}\partial_{t}\partial_{x}u_{\varepsilon}
-\partial_{x}^{2}u_{\varepsilon}\partial_{t}\partial_{x}^{2}u_{\varepsilon}]\\[5pt]
&=\partial_{x}^{3}A_{\mu}^{-1}[2\mu(u_{\varepsilon})\partial_{t}u_{\varepsilon}
+\partial_{x}u_{\varepsilon}\partial_{t}\partial_{x}u_{\varepsilon}
-\frac{1}{2}\partial_{x}u_{\varepsilon}(\partial_{x}^{2}u_{\varepsilon})^{2}
+\partial_{x}^{2}u_{\varepsilon}\partial_{x}^{3}P_{1,\varepsilon}\\[5pt]
&\qquad\qquad
+3\partial_{x}^{2}u_{\varepsilon}\mu(A_{\mu}^{-1}(\partial_{x}u_{\varepsilon}\partial_{x}^{2}u_{\varepsilon}))
-3\partial_{x}^{2}u_{\varepsilon}\partial_{x}^{2}A_{\mu}^{-1}(\partial_{x}u_{\varepsilon}\partial_{x}^{2}u_{\varepsilon})
+\varepsilon(\partial_{x}^{3}u_{\varepsilon})^{2}]\\[5pt]
&\quad+\partial_{x}^{4}A_{\mu}^{-1}[\frac{1}{2}u_{\varepsilon}(\partial_{x}^{2}u_{\varepsilon})^{2}
-\varepsilon\partial_{x}^{2}u_{\varepsilon}\partial_{x}^{3}u_{\varepsilon}]\\[5pt]
&:=\partial_{x}^{3}A_{\mu}^{-1}(E_{1})
+\partial_{x}^{4}A_{\mu}^{-1}(E_{2}).
\end{array}
$$
Note that
$$
\begin{array}{rl}
\partial_{x}^{4}A_{\mu}^{-1}(E_{2})
=-\mu(A_{\mu}^{-1}(E_{2}))
+\partial_{x}^{2}A_{\mu}^{-1}(E_{2})
+E_{2}.
\end{array}
$$
By Lemmas 5.1-5.2, (5.6), (5.7) and Young's inequality, we get
$$
\begin{array}{rl}
\int_{0}^{T}\int_{\mathbb{S}}|\partial_{t}\partial_{x}^{3}P_{1,\varepsilon}|dxdt
&=\int_{0}^{T}\int_{\mathbb{S}}|\partial_{x}^{3}g*E_{1}
-\mu(g*E_{2})+\partial_{x}^{2}g*E_{2}+E_{2}|dxdt\\[5pt]
&\leq C_{0}(\|E_{1}\|_{L^{1}([0, T]\times\mathbb{S})}
 +\|E_{2}\|_{L^{1}([0, T]\times\mathbb{S})})\\[5pt]
&\leq C_{0}(T+1)(\|u_{0}\|_{H^{2}(\mathbb{S})}^{3}+\|u_{0}\|_{H^{2}(\mathbb{S})}^{2}),
\end{array}
$$
which completes the proof of Lemma 5.3. \hfill $\Box$\\

\noindent\textbf{Lemma 5.4.} Let $u_{0}\in H^{2}(\mathbb{S})$ and $\partial_{x}^{2}u_{0}\in L^{p}(\mathbb{S})$ for some $p>2$. Then the following inequality holds
$$
\begin{array}{rl}
\|\partial_{x}^{2}u_{\varepsilon}(t, \cdot)\|_{L^{p}(\mathbb{S})}
\leq e^{\|u_{0}\|_{H^{2}(\mathbb{S})}t}\|\partial_{x}^{2}u_{0}\|_{L^{p}(\mathbb{S})}
+pC_{0}\|u_{0}\|_{H^{2}(\mathbb{S})}(e^{\|u_{0}\|_{H^{2}(\mathbb{S})}t}-1).
\end{array}
$$

\noindent\textbf{Proof.}
Denote $q_{\varepsilon}:=\partial_{x}^{2}u_{\varepsilon}$, then $q_{\varepsilon}$ satisfies
$$
\begin{array}{rl}
\partial_{t}q_{\varepsilon}
+u_{\varepsilon}\partial_{x}q_{\varepsilon}
+\partial_{x}^{3}P_{1,\varepsilon}
+3\mu(A_{\mu}^{-1}(q_{\varepsilon}\partial_{x}u_{\varepsilon}))
-3\partial_{x}^{2}A_{\mu}^{-1}(q_{\varepsilon}\partial_{x}u_{\varepsilon})
=\varepsilon\partial_{x}^{2}q_{\varepsilon}.
\end{array}
\eqno(5.8)
$$
Multiplying the above equation by $pq_{\varepsilon}|q_{\varepsilon}|^{p-2}$, we have
$$
\begin{array}{rl}
&\partial_{t}(|q_{\varepsilon}|^{p})
+u_{\varepsilon}\partial_{x}(|q_{\varepsilon}|^{p})
+pq_{\varepsilon}|q_{\varepsilon}|^{p-2}\partial_{x}^{3}P_{1,\varepsilon}
+3pq_{\varepsilon}|q_{\varepsilon}|^{p-2}\mu(A_{\mu}^{-1}(q_{\varepsilon}\partial_{x}u_{\varepsilon}))\\[5pt]
&-3pq_{\varepsilon}|q_{\varepsilon}|^{p-2}\partial_{x}^{2}A_{\mu}^{-1}(q_{\varepsilon}\partial_{x}u_{\varepsilon})
=\varepsilon pq_{\varepsilon}|q_{\varepsilon}|^{p-2}\partial_{x}^{2}q_{\varepsilon}
=\varepsilon\partial_{x}^{2}(|q_{\varepsilon}|^{p})
-\varepsilon p(p-1)|q_{\varepsilon}|^{p-2}(\partial_{x}q_{\varepsilon})^{2}.
\end{array}
$$
By Lemmas 5.1-5.2 and Young's inequality, we know
$$
\begin{array}{rl}
&\frac{d}{dt}\int_{\mathbb{S}}|q_{\varepsilon}|^{p}dx\\[5pt]
&\leq \int_{\mathbb{S}}|q_{\varepsilon}|^{p}\partial_{x}u_{\varepsilon}dx
+p\int_{\mathbb{S}}|q_{\varepsilon}|^{p-1}|\partial_{x}^{3}P_{1,\varepsilon}|dx\\[5pt]
&\quad+3p\int_{\mathbb{S}}|q_{\varepsilon}|^{p-1}|\mu(A_{\mu}^{-1}(q_{\varepsilon}\partial_{x}u_{\varepsilon}))|dx
+3p\int_{\mathbb{S}}|q_{\varepsilon}|^{p-1}|\partial_{x}^{2}A_{\mu}^{-1}(q_{\varepsilon}\partial_{x}u_{\varepsilon})|dx\\[5pt]
&\leq \|\partial_{x}u_{\varepsilon}\|_{L^{\infty}(\mathbb{S})}
\int_{\mathbb{S}}|q_{\varepsilon}|^{p}dx
+p\|q_{\varepsilon}\|_{L^{p}(\mathbb{S})}^{p-1}
\|\partial_{x}^{3}P_{1,\varepsilon}\|_{L^{p}(\mathbb{S})}\\[5pt]
&\quad
+3p\|q_{\varepsilon}\|_{L^{p}(\mathbb{S})}^{p-1}
\|A_{\mu}^{-1}(q_{\varepsilon}\partial_{x}u_{\varepsilon})\|_{L^{1}(\mathbb{S})}
+3p\|q_{\varepsilon}\|_{L^{p}(\mathbb{S})}^{p-1}
\|\partial_{x}^{2}A_{\mu}^{-1}(q_{\varepsilon}\partial_{x}u_{\varepsilon})\|_{L^{p}(\mathbb{S})}\\[5pt]
&\leq \|\partial_{x}u_{\varepsilon}\|_{L^{\infty}(\mathbb{S})}
\int_{\mathbb{S}}|q_{\varepsilon}|^{p}dx
+p\|q_{\varepsilon}\|_{L^{p}(\mathbb{S})}^{p-1}
\|\partial_{x}^{3}P_{1,\varepsilon}\|_{L^{p}(\mathbb{S})}\\[5pt]
&\quad
+3p\|q_{\varepsilon}\|_{L^{p}(\mathbb{S})}^{p-1}
\|g\|_{L^{1}(\mathbb{S})}\|q_{\varepsilon}\|_{L^{2}(\mathbb{S})}
\|\partial_{x}u_{\varepsilon}\|_{L^{\infty}(\mathbb{S})}
+3p\|q_{\varepsilon}\|_{L^{p}(\mathbb{S})}^{p-1}
\|\partial_{x}^{2}g\|_{L^{p}(\mathbb{S})}
\|q_{\varepsilon}\|_{L^{2}(\mathbb{S})}
\|\partial_{x}u_{\varepsilon})\|_{L^{\infty}(\mathbb{S})}\\[5pt]
&\leq \|u_{0}\|_{H^{2}(\mathbb{S})}\int_{\mathbb{S}}|q_{\varepsilon}|^{p}dx
+pC_{0}\|u_{0}\|_{H^{2}(\mathbb{S})}^{2}\|q_{\varepsilon}\|_{L^{p}(\mathbb{S})}^{p-1}.
\end{array}
$$
Note that
$$
\begin{array}{rl}
\frac{d}{dt}\int_{\mathbb{S}}|q_{\varepsilon}|^{p}dx
=p\|q_{\varepsilon}\|_{L^{p}(\mathbb{S})}^{p-1}\frac{d}{dt}\|q_{\varepsilon}\|_{L^{p}(\mathbb{S})}.
\end{array}
$$
Thus,
$$
\begin{array}{rl}
\frac{d}{dt}\|q_{\varepsilon}\|_{L^{p}(\mathbb{S})}
\leq \|u_{0}\|_{H^{2}(\mathbb{S})}\|q_{\varepsilon}\|_{L^{p}(\mathbb{S})}
+pC_{0}\|u_{0}\|_{H^{2}(\mathbb{S})}^{2}.
\end{array}
$$
The Gronwall inequality implies the desired result.
\hfill $\Box$\\

To convenient, we define
$$
\begin{array}{rl}
Q_{\varepsilon}
&:=3\partial_{x}u_{\varepsilon}\partial_{x}^{2}u_{\varepsilon}
+\partial_{x}^{3}P_{\varepsilon}\\[5pt]
&=\partial_{x}^{3}P_{1,\varepsilon}
+3\mu(A_{\mu}^{-1}(\partial_{x}u_{\varepsilon}\partial_{x}^{2}u_{\varepsilon}))
-3\partial_{x}^{2}A_{\mu}^{-1}(\partial_{x}u_{\varepsilon}\partial_{x}^{2}u_{\varepsilon}).
\end{array}
$$

\noindent\textbf{Lemma 5.5.} Let $u_{0}\in H^{2}(\mathbb{S})$ and $\partial_{x}^{2}u_{0}\in L^{p}(\mathbb{S})$ for some $p>2$. There exist a positive sequence $\{\varepsilon_{k}\}_{k\in \mathbb{N}}$ decreasing to zero and three functions $u\in L^{\infty}(\mathbb{R}_{+}; H^{2}(\mathbb{S}))\cap H^{1}(\mathbb{R}_{+}\times\mathbb{S})\subseteq C(\mathbb{R}_{+}; C^{1}(\mathbb{S}))$ for each $T>0$, $P\in L^{\infty}(\mathbb{R}_{+}; W^{2, \infty}(\mathbb{S}))$
and $Q\in L^{\infty}(\mathbb{R}_{+}; W^{1, 1}(\mathbb{S})\cap L^{\infty}(\mathbb{S}))$ such that
$$
\begin{array}{rl}
&u_{\varepsilon_{k}}\rightharpoonup u \quad
\mbox{weakly~in}~H^{1}([0, T]\times \mathbb{S})~\mbox{for~each}~T\geq 0;\\[5pt]
&u_{\varepsilon_{k}}\rightarrow u \quad
\mbox{strongly~in}~L_{loc}^{\infty}(\mathbb{R}_{+}; H^{1}(\mathbb{S}));\\[5pt]
&P_{\varepsilon_{k}}\rightharpoonup P \quad
\mbox{weakly~in}~L_{loc}^{\sigma}(\mathbb{R}_{+}\times\mathbb{S})~\mbox{for~each}~1< \sigma<\infty;\\[5pt]
&Q_{\varepsilon_{k}}\rightarrow Q \quad
\mbox{strongly~in}~L_{loc}^{\sigma}(\mathbb{R}_{+}\times\mathbb{S})~\mbox{for~each}~1\leq \sigma<\infty.
\end{array}
$$

\noindent\textbf{Proof.}
Due to Lemmas 5.1 and 5.3, we have that
$$
\begin{array}{c}
\{u_{\varepsilon}\}_{\varepsilon}~
\mbox{is~uniformly~bounded~in}~L^{\infty}(\mathbb{R}_{+}; H^{2}(\mathbb{S})),\\[5pt]
\{\partial_{t}u_{\varepsilon}\}_{\varepsilon}~
\mbox{is~uniformly~bounded~in}~L^{2}([0, T]; H^{1}(\mathbb{S}))~\mbox{for~each}~T> 0.
\end{array}
$$
In particular, $\{u_{\varepsilon}\}_{\varepsilon}$ is uniformly bounded in
$H^{1}([0, T]\times\mathbb{S})$ and then we have $u_{\varepsilon_{k}}\rightharpoonup u$
weakly in $H^{1}([0, T]\times \mathbb{S})$.
Moreover, Using the fact $H^{2}(\mathbb{S})\Subset H^{1}(\mathbb{S})\Subset L^{2}(\mathbb{S})$ and Corollary 4 in \cite{simon87}, we know that $u_{\varepsilon_{k}}\rightarrow u$ strongly in $L_{loc}^{\infty}(\mathbb{R}_{+}; H^{1}(\mathbb{S}))$.

Due to Lemma 5.2, we obtain that $\{P_{\varepsilon}\}_{\varepsilon}$
is uniformly bounded in $L^{\infty}(\mathbb{R}_{+}; W^{2, \infty}(\mathbb{S}))$.
In particular, $\{P_{\varepsilon}\}_{\varepsilon}$ is uniformly bounded in
$L^{\sigma}([0, T]\times\mathbb{S})$ with $1< \sigma<\infty$ and then we have
$P_{\varepsilon_{k}}\rightharpoonup P$ weakly in $L_{loc}^{\sigma}(\mathbb{R}_{+}\times\mathbb{S})$ for each $1< \sigma<\infty$.

Moreover, due to Lemmas 5.2-5.3, we know
$$
\begin{array}{rl}
\|Q_{\varepsilon}\|_{L^{\infty}(\mathbb{S})}
&=\|\partial_{x}^{3}P_{1,\varepsilon}
+3\mu(A_{\mu}^{-1}(\partial_{x}u_{\varepsilon}\partial_{x}^{2}u_{\varepsilon}))
+\partial_{x}P_{2,\varepsilon}\|_{L^{\infty}(\mathbb{S})}\\[5pt]
&\leq \|\partial_{x}^{3}P_{1,\varepsilon}\|_{L^{\infty}(\mathbb{S})}
+3\|u_{\varepsilon}\partial_{x}^{2}u_{\varepsilon}\|_{L^{1}(\mathbb{S})}
+\|\partial_{x}P_{2,\varepsilon}\|_{L^{\infty}(\mathbb{S})}\\[5pt]
&\leq C_{0}\|u_{0}\|_{H^{2}(\mathbb{S})}^{2}
\end{array}
$$
and
$$
\begin{array}{rl}
\|\partial_{x}Q_{\varepsilon}\|_{L^{1}(\mathbb{S})}
&=\|\partial_{x}^{4}P_{1,\varepsilon}
+\partial_{x}^{2}P_{2,\varepsilon}\|_{L^{1}(\mathbb{S})}\\[5pt]
&\leq \|\partial_{x}^{4}P_{1,\varepsilon}\|_{L^{1}(\mathbb{S})}
+\|\partial_{x}^{2}P_{2,\varepsilon}\|_{L^{1}(\mathbb{S})}\\[5pt]
&\leq C_{0}\|u_{0}\|_{H^{2}(\mathbb{S})}^{2},
\end{array}
$$
then $\{Q_{\varepsilon}\}_{\varepsilon}$ is uniformly bounded in $L^{\infty}(\mathbb{R}_{+}; W^{1,1}(\mathbb{S})\cap L^{\infty}(\mathbb{S}))$.
Note that
$$
\begin{array}{rl}
&\partial_{x}u_{\varepsilon}\partial_{t}\partial_{x}^{2}u_{\varepsilon}\\[5pt]
&=\partial_{x}u_{\varepsilon}(u_{\varepsilon}\partial_{x}^{3}u_{\varepsilon}
+3\partial_{x}u_{\varepsilon}\partial_{x}^{2}u_{\varepsilon}
+\partial_{x}^{3}P_{\varepsilon}
-\varepsilon\partial_{x}^{4}u_{\varepsilon})\\[5pt]
&=\partial_{x}(u_{\varepsilon}\partial_{x}u_{\varepsilon}\partial_{x}^{2}u_{\varepsilon})
-u_{\varepsilon}(\partial_{x}^{2}u_{\varepsilon})^{2}
-(\partial_{x}u_{\varepsilon})^{2}\partial_{x}^{2}u_{\varepsilon}
+\partial_{x}u_{\varepsilon}\partial_{x}^{3}P_{1,\varepsilon}\\[5pt]
&\quad+3\partial_{x}u_{\varepsilon}\mu(A_{\mu}^{-1}(\partial_{x}u_{\varepsilon}\partial_{x}^{2}u_{\varepsilon}))
-3\partial_{x}u_{\varepsilon}\partial_{x}^{2}A_{\mu}^{-1}(\partial_{x}u_{\varepsilon}\partial_{x}^{2}u_{\varepsilon})
-\varepsilon\partial_{x}(\partial_{x}u_{\varepsilon}\partial_{x}^{3}u_{\varepsilon})
+\varepsilon\partial_{x}^{2}u_{\varepsilon}\partial_{x}^{3}u_{\varepsilon}.
\end{array}
$$
By Young's inequality and Lemma 5.1,
$$
\begin{array}{rl}
&\|\partial_{t}Q_{\varepsilon}\|_{L^{1}([0, T]\times\mathbb{S})}\\[5pt]
&=\|\partial_{t}\partial_{x}^{3}P_{1,\varepsilon}
+3\mu(A_{\mu}^{-1}\partial_{t}(\partial_{x}u_{\varepsilon}\partial_{x}^{2}u_{\varepsilon}))
-3\partial_{x}^{2}A_{\mu}^{-1}\partial_{t}(\partial_{x}u_{\varepsilon}\partial_{x}^{2}u_{\varepsilon})\|_{L^{1}([0, T]\times\mathbb{S})}\\[5pt]
&\lesssim\|\partial_{t}\partial_{x}^{3}P_{1,\varepsilon}\|_{L^{1}([0, T]\times\mathbb{S})}
+\|\partial_{t}\partial_{x}u_{\varepsilon}\partial_{x}^{2}u_{\varepsilon}\|_{L^{1}([0, T]\times\mathbb{S})}\\[5pt]
&\quad
+\|A_{\mu}^{-1}(\partial_{x}u_{\varepsilon}\partial_{t}\partial_{x}^{2}u_{\varepsilon})\|_{L^{1}([0, T]\times\mathbb{S})}
+\|\partial_{x}^{2}A_{\mu}^{-1}(\partial_{x}u_{\varepsilon}\partial_{t}\partial_{x}^{2}u_{\varepsilon})\|_{L^{1}([0, T]\times\mathbb{S})}\\[5pt]
&\lesssim\|\partial_{t}\partial_{x}^{3}P_{1,\varepsilon}\|_{L^{1}([0, T]\times\mathbb{S})}
+\|\partial_{t}\partial_{x}u_{\varepsilon}\|_{L^{2}([0, T]\times\mathbb{S})}
\|\partial_{x}^{2}u_{\varepsilon}\|_{L^{2}([0, T]\times\mathbb{S})}\\[5pt]
&\quad+\|u_{\varepsilon}\partial_{x}u_{\varepsilon}\partial_{x}^{2}u_{\varepsilon}\|_{L^{1}([0, T]\times\mathbb{S})}
+\|u_{\varepsilon}(\partial_{x}^{2}u_{\varepsilon})^{2}\|_{L^{1}([0, T]\times\mathbb{S})}
+\|(\partial_{x}u_{\varepsilon})^{2}\partial_{x}^{2}u_{\varepsilon}\|_{L^{1}([0, T]\times\mathbb{S})}\\[5pt]
&\quad+\|\partial_{x}u_{\varepsilon}\partial_{x}^{3}P_{1,\varepsilon}\|_{L^{1}([0, T]\times\mathbb{S})}
+\|\partial_{x}u_{\varepsilon}\|_{L^{\infty}([0, T]\times\mathbb{S})}
\|\partial_{x}u_{\varepsilon}\partial_{x}^{2}u_{\varepsilon}\|_{L^{1}([0, T]\times\mathbb{S})}\\[5pt]
&\quad+\varepsilon\|\partial_{x}u_{\varepsilon}\partial_{x}^{3}u_{\varepsilon}\|_{L^{1}([0, T]\times\mathbb{S})}
+\varepsilon\|\partial_{x}^{2}u_{\varepsilon}\partial_{x}^{3}u_{\varepsilon}\|_{L^{1}([0, T]\times\mathbb{S})}\\[5pt]
&\leq C_{0}(T+1)(\|u_{0}\|_{H^{2}(\mathbb{S})}^{3}+\|u_{0}\|_{H^{2}(\mathbb{S})}^{2}),
\end{array}
$$
which implies that $\{\partial_{t}Q_{\varepsilon}\}_{\varepsilon}$ is uniformly bounded in $L^{1}([0, T]\times\mathbb{S})$ for each $T>0$.
Using the fact $W^{1, 1}(\mathbb{S})\Subset L^{\sigma}(\mathbb{S})\subset L^{1}(\mathbb{S})$, $1\leq\sigma<\infty$ and Corollary 4 in \cite{simon87},
we know that $Q_{\varepsilon_{k}}\rightarrow Q$
strongly in $L_{loc}^{\sigma}(\mathbb{R}_{+}\times\mathbb{S})$ for each $1\leq\sigma<\infty$. \hfill $\Box$

\subsection{\textbf{Existence of solutions}}

From Lemmas 5.1 and 5.4, we can deduce that there exist two functions
$q\in L_{loc}^{\rho}(\mathbb{R}_{+}\times\mathbb{S})$
and $\overline{q^{2}}\in L_{loc}^{r}(\mathbb{R}_{+}\times\mathbb{S})$ such that
$$
\begin{array}{l}
q_{\varepsilon_{k}}\rightharpoonup q \quad
\mbox{in}~L_{loc}^{\rho}(\mathbb{R}_{+}\times\mathbb{S}),
\quad
q^{2}_{\varepsilon_{k}}\rightharpoonup \overline{q^{2}} \quad
\mbox{in}~L_{loc}^{r}(\mathbb{R}_{+}\times\mathbb{S}),
\end{array}
\eqno(5.9)
$$
for every $1<\rho<p$ and $1<r<\frac{p}{2}$. Moreover,
$$
q^{2}(t,x)\leq \overline{q^{2}}(t,x),\quad a.e.~(t,x)\in \mathbb{R}_{+}\times\mathbb{S}.
$$

In view of (5.9), we conclude that for any $\eta\in C^{1}(\mathbb{R})$ with $\eta^{\prime}$
bounded, Lipschitz continuous on $\mathbb{R}$, $\eta(0)=0$ and any $1<\rho<p$, we have
$$
\eta(q_{\varepsilon_{k}})\rightharpoonup \overline{\eta(q)}
\quad \mbox{in}~L_{loc}^{\rho}(\mathbb{R}_{+}\times\mathbb{S}).
$$
Here and in what follows, we use overbars to denote weak limits in spaces to be understood
from the context.\\

\noindent\textbf{Lemma 5.6.}
The following inequality holds in the sense of distributions
$$
\begin{array}{rl}
&\int_{\mathbb{S}}\left(\overline{(q_{+})^{2}}-(q_{+})^{2}\right)dx
\leq \int_{0}^{t}\int_{\mathbb{S}}\partial_{x}u\left(\overline{(q_{+})^{2}}-(q_{+})^{2}\right)dtdx
-2\int_{0}^{t}\int_{\mathbb{S}}Q(\overline{q_{+}}-q_{+})dtdx.
\end{array}
$$
\noindent\textbf{Proof.}
Taking $\xi\in C^{2}(\mathbb{R})$ convex
and multiplying (5.8) by $\xi^{\prime}(q_{\varepsilon_{k}})$, we have
$$
\begin{array}{rl}
&\partial_{t}\xi(q_{\varepsilon_{k}})
+\partial_{x}(u_{\varepsilon_{k}}\xi(q_{\varepsilon_{k}}))
-\xi(q_{\varepsilon_{k}})\partial_{x}u_{\varepsilon_{k}}
+\xi^{\prime}(q_{\varepsilon_{k}})Q_{\varepsilon_{k}}\\[5pt]
&=\varepsilon_{k}\partial_{x}^{2}\xi(q_{\varepsilon_{k}})
-\varepsilon_{k}\xi^{\prime\prime}(q_{\varepsilon_{k}})(\partial_{x}q_{\varepsilon_{k}})^{2}
\leq\varepsilon_{k}\partial_{x}^{2}\xi(q_{\varepsilon_{k}}).
\end{array}
$$
In particular, we can use the entropy $q\mapsto (q_{+})^{2}/2$ and get
$$
\begin{array}{rl}
&\partial_{t}\frac{(q_{\varepsilon_{k},+})^{2}}{2}
+\partial_{x}
\left(u_{\varepsilon_{k}}\frac{(q_{\varepsilon_{k},+})^{2}}{2}\right)
-\frac{(q_{\varepsilon_{k},+})^{2}}{2}\partial_{x}u_{\varepsilon_{k}}
+q_{\varepsilon_{k},+}Q_{\varepsilon_{k}}
\leq \varepsilon_{k}\partial_{x}^{2}\frac{(q_{\varepsilon_{k},+})^{2}}{2}.
\end{array}
$$
Letting $k\rightarrow \infty$, we have
$$
\begin{array}{l}
\partial_{t}\frac{\overline{(q_{+})^{2}}}{2}
+\partial_{x}
\left(u\frac{\overline{(q_{+})^{2}}}{2}\right)
-\frac{\overline{(q_{+})^{2}}}{2}\partial_{x}u
+Q\overline{q_{+}}\leq 0.
\end{array}
$$
The rest of the proof is the same as in \cite{wang17}.
\hfill $\Box$\\

Similar as the proof of Lemmas 5.7-5.8 in \cite{wang17}, we can obtain the following two lemmas by using the entropy $\eta_{R}^{-}(\xi):=\eta_{R}(\xi)\chi_{[-\infty,0]}(\xi)$,
where $R>0$, $\chi_{E}$ is the characteristic function in set $E$ and
$$
\begin{array}{rl}
\eta_{R}(\xi)
:=\left\{\begin{array}{l}
\frac{1}{2}\xi^{2}, \quad \mbox{if}~|\xi|\leq R, \\[5pt]
R|\xi|-\frac{1}{2}R^{2}, \quad \mbox{if}~|\xi|>R.
\end{array}
\right.
\end{array}
$$.

\noindent\textbf{Lemma 5.7.} For any $t>0$ and any $R>0$,
$$
\begin{array}{rl}
&\int_{\mathbb{S}}\left(\overline{\eta_{R}^{-}(q)}-\eta_{R}^{-}(q)\right)dx
\leq
\int_{0}^{t}\int_{\mathbb{S}}\partial_{x}u\left(\overline{\eta_{R}^{-}(q)}-\eta_{R}^{-}(q)\right)dtdx\\[5pt]
&\quad-\int_{0}^{t}\int_{\mathbb{S}}Q
\left(\overline{(\eta_{R}^{-})^{\prime}(q)}-(\eta_{R}^{-})^{\prime}(q)\right)dtdx. \end{array}
$$

\noindent\textbf{Lemma 5.8.} There holds $\overline{q^{2}}=q^{2}$, a.e. on $\mathbb{R}_{+}\times \mathbb{S}$.\\

\noindent\textbf{Proof of Theorem 5.1.}
Let $u(t,x)$ be the limit of the viscous approximation solutions $u_{\varepsilon}(t,x)$
as $\varepsilon\rightarrow 0$. It then follows from Lemmas 5.1 and 5.5 that
$u(t,x)\in C(\mathbb{R}_{+}; C^{1}(\mathbb{S}))\cap L^{\infty}(\mathbb{R}_{+}; H^{2}(\mathbb{S}))$
and (5.2) holds.

Let $\mu_{t,x}(\lambda)$ be the Young measure
associated with $\{q_{\varepsilon}\}=\{\partial_{x}^{2}u_{\varepsilon}\}$, see more details in \cite{xz00}.
By Lemma 5.8, we have $\mu_{t,x}(\lambda)=\delta_{\overline{q(t,x)}}(\lambda)$ a.e.
$(t,x)\in \mathbb{R}_{+}\times \mathbb{S}$, then
$$
\begin{array}{l}
q_{\varepsilon}=\partial_{x}^{2}u_{\varepsilon}\rightarrow q=\partial_{x}^{2}u
\quad\mbox{in}~L^{2}_{loc}(\mathbb{R}_{+}\times \mathbb{S}).
\end{array}
\eqno(5.10)
$$
Taking $\varepsilon\rightarrow 0$ in Eq. (5.3), one finds from (5.10)
and Lemma 5.5 that $u(t,x)$ is an admissible weak solution to Eq. (5.1).
This completes the proof of Theorem 5.1. \hfill $\Box$

\section{Peaked solutions}

In this section, we show the existence of single peakon solutions to Eq.(1.4).\\

\noindent\textbf{Theorem 6.1.} For any $c>0$, Eq.(1.4) admits the peaked periodic-one traveling-wave solutions $u_{c}(t,x)=\phi(\xi)$, $\xi=x-ct$, where
$$
\begin{array}{l}
\phi(\xi)
=\frac{12\sinh(\frac{1}{2})}{25\sinh(\frac{1}{2})-6\cosh(\frac{1}{2})}c
\left[\frac{1}{2}(\xi-[\xi]-\frac{1}{2})^{2}
-\frac{\cosh(\xi-[\xi]-\frac{1}{2})}{2\sinh(\frac{1}{2})}
+\frac{47}{24}\right].
\end{array}
$$

\noindent\textbf{Proof.} Motivated by the forms of periodic peakons for the $\mu$-Camassa-Holm equation \cite{lmt10}, we assume that the periodic peakon of (1.4) is given by
$$
\begin{array}{l}
u_{c}(t, x)
=a\left[\frac{1}{2}(\xi-[\xi]-\frac{1}{2})^{2}
-\frac{\cosh(\xi-[\xi]-\frac{1}{2})}{2\sinh(\frac{1}{2})}
+\frac{47}{24}\right].
\end{array}
$$
According to the definition of weak solutions, $u_{c}(t, x)$ satisfies the following equation
$$
\begin{array}{rl}
\Sigma_{j=1}^{4}I_{j}
&:=\int_{0}^{T}\int_{\mathbb{S}}u_{c,t}\varphi dxdt
+\int_{0}^{T}\int_{\mathbb{S}}u_{c}u_{c,x}\varphi dxdt\\[5pt]
&\quad+\int_{0}^{T}\int_{\mathbb{S}}g_{x}*[2\mu(u_{c})u_{c}+\frac{1}{2}u_{c,x}^{2}-\frac{1}{2}u_{c,xx}^{2}]\varphi dxdt
-\int_{0}^{T}\int_{\mathbb{S}}3g_{xx}*(u_{c,x}u_{c,xx})\varphi dxdt\\[5pt]
&=0,
\end{array}
$$
for some $T>0$ and any test function $\varphi(t,x)\in C_{c}^{\infty}([0, T)\times \mathbb{S})$,
where $g(x)=\frac{1}{2}(x-[x]-\frac{1}{2})^{2}
-\frac{\cosh(x-[x]-\frac{1}{2})}{2\sinh(\frac{1}{2})}
+\frac{47}{24}$. One can obtain that
$$
\begin{array}{rl}
\mu(u_{c})
&=a\int_{0}^{ct}\left[\frac{1}{2}(\xi+\frac{1}{2})^{2}
-\frac{\cosh(\xi+\frac{1}{2})}{2\sinh(\frac{1}{2})}
+\frac{47}{24}\right]dx\\[5pt]
&\quad+a\int_{ct}^{1}\left[\frac{1}{2}(\xi-\frac{1}{2})^{2}
-\frac{\cosh(\xi-\frac{1}{2})}{2\sinh(\frac{1}{2})}
+\frac{47}{24}\right]dx\\[5pt]
&=a.
\end{array}
$$

To evaluate $I_{j}$, $j=1,2,3,4$, we need to consider two cases: $(i)$ $x>ct$
and $x\leq ct$.

For $x>ct$, we get
$$
\begin{array}{rl}
&g_{x}*[2\mu(u_{c})u_{c}+\frac{1}{2}u_{c,x}^{2}-\frac{1}{2}u_{c,xx}^{2}]\\[5pt]
&=a^{2}\int_{\mathbb{S}}\left((x-y-[x-y]-\frac{1}{2})
-\frac{\sinh(x-y-[x-y]-\frac{1}{2})}{2\sinh(\frac{1}{2})}\right)\\[5pt]
&\quad \times((y-ct-[y-ct]-\frac{1}{2})^{2}
-\frac{\cosh(y-ct-[y-ct]-\frac{1}{2})}{\sinh(\frac{1}{2})}+\frac{47}{12}\\[5pt]
&\quad+\frac{1}{2}(y-ct-[y-ct]-\frac{1}{2}-\frac{\sinh(y-ct-[y-ct]-\frac{1}{2})}{2\sinh(\frac{1}{2})})^{2}
-\frac{1}{2}(1-\frac{\cosh(y-ct-[y-ct]-\frac{1}{2})}{\sinh(\frac{1}{2})})^{2})dy\\[5pt]
&=a^{2}\int_{0}^{ct}\left((x-y-\frac{1}{2})
-\frac{\sinh(x-y-\frac{1}{2})}{2\sinh(\frac{1}{2})}\right)\\[5pt]
&\quad \times\left((y-ct+\frac{1}{2})^{2}
-\frac{\cosh(y-ct+\frac{1}{2})}{\sinh(\frac{1}{2})}+\frac{47}{12}
+\frac{1}{2}(y-ct+\frac{1}{2}-\frac{\sinh(y-ct+\frac{1}{2})}{2\sinh(\frac{1}{2})})^{2}
-\frac{1}{2}(1-\frac{\cosh(y-ct+\frac{1}{2})}{\sinh(\frac{1}{2})})^{2}\right)dy\\[5pt]
&\quad+a^{2}\int_{ct}^{x}\left((x-y-\frac{1}{2})
-\frac{\sinh(x-y-\frac{1}{2})}{2\sinh(\frac{1}{2})}\right)\\[5pt]
&\quad \times\left((y-ct-\frac{1}{2})^{2}
-\frac{\cosh(y-ct-\frac{1}{2})}{\sinh(\frac{1}{2})}+\frac{47}{12}
+\frac{1}{2}(y-ct-\frac{1}{2}-\frac{\sinh(y-ct-\frac{1}{2})}{2\sinh(\frac{1}{2})})^{2}
-\frac{1}{2}(1-\frac{\cosh(y-ct-\frac{1}{2})}{\sinh(\frac{1}{2})})^{2}\right)dy\\[5pt]
&\quad+a^{2}\int_{x}^{1}\left((x-y+\frac{1}{2})
-\frac{\sinh(x-y+\frac{1}{2})}{2\sinh(\frac{1}{2})}\right)\\[5pt]
&\quad \times\left((y-ct-\frac{1}{2})^{2}
-\frac{\cosh(y-ct-\frac{1}{2})}{\sinh(\frac{1}{2})}+\frac{47}{12}
+\frac{1}{2}(y-ct-\frac{1}{2}-\frac{\sinh(y-ct-\frac{1}{2})}{2\sinh(\frac{1}{2})})^{2}
-\frac{1}{2}(1-\frac{\cosh(y-ct-\frac{1}{2})}{\sinh(\frac{1}{2})})^{2}\right)dy
\end{array}
$$
and
$$
\begin{array}{rl}
&g_{xx}*(u_{c,x}u_{c,xx})\\[5pt]
&=a^{2}\int_{\mathbb{S}}\left(1-\frac{\cosh(x-y-[x-y]-\frac{1}{2})}{2\sinh(\frac{1}{2})}\right)\\[5pt]
&\quad \times\left((y-ct-[y-ct]-\frac{1}{2})
-\frac{\sinh(y-ct-[y-ct]-\frac{1}{2})}{2\sinh(\frac{1}{2})}\right)
\left(1-\frac{\cosh(y-ct-[y-ct]-\frac{1}{2})}{2\sinh(\frac{1}{2})}\right)dy\\[5pt]
&=a^{2}\int_{0}^{ct}\left(1-\frac{\cosh(x-y-\frac{1}{2})}{2\sinh(\frac{1}{2})}\right)
\left((y-ct+\frac{1}{2})
-\frac{\sinh(y-ct+\frac{1}{2})}{2\sinh(\frac{1}{2})}\right)
\left(1-\frac{\cosh(y-ct+\frac{1}{2})}{2\sinh(\frac{1}{2})}\right)dy\\[5pt]
&\quad+a^{2}\int_{ct}^{x}\left(1-\frac{\cosh(x-y-\frac{1}{2})}{2\sinh(\frac{1}{2})}\right)
\left((y-ct-\frac{1}{2})
-\frac{\sinh(y-ct-\frac{1}{2})}{2\sinh(\frac{1}{2})}\right)
\left(1-\frac{\cosh(y-ct-\frac{1}{2})}{2\sinh(\frac{1}{2})}\right)dy\\[5pt]
&\quad+a^{2}\int_{x}^{1}\left(1-\frac{\cosh(x-y+\frac{1}{2})}{2\sinh(\frac{1}{2})}\right)
\left((y-ct-\frac{1}{2})
-\frac{\sinh(y-ct-\frac{1}{2})}{2\sinh(\frac{1}{2})}\right)
\left(1-\frac{\cosh(y-ct-\frac{1}{2})}{2\sinh(\frac{1}{2})}\right)dy.
\end{array}
$$
By calculations, we have
$$
\begin{array}{rl}
&g_{x}*[2\mu(u_{c})u_{c}+\frac{1}{2}u_{c,x}^{2}-\frac{1}{2}u_{c,xx}^{2}]
-3g_{xx}*(u_{c,x}u_{c,xx})\\[5pt]
&=a^{2}[-\frac{1}{2}(x-ct-\frac{1}{2})^{3}
+(x-ct-\frac{1}{2})(\frac{1}{8}-\frac{\cosh(\frac{1}{2})}{2\sinh(\frac{1}{2})}
+\frac{\cosh(x-ct-\frac{1}{2})}{2\sinh(\frac{1}{2})})\\[5pt]
&\quad\quad +(x-ct)^{2}\frac{\sinh(x-ct-\frac{1}{2})}{4\sinh(\frac{1}{2})}
-(x-ct)\frac{\sinh(x-ct-\frac{1}{2})}{4\sinh(\frac{1}{2})}\\[5pt]
&\quad\quad -\frac{\sinh(x-ct-\frac{1}{2})\cosh(x-ct-\frac{1}{2})}{4(\sinh(\frac{1}{2}))^{2}}
+\frac{\sinh(x-ct-\frac{1}{2})\cosh(\frac{1}{2})}{4(\sinh(\frac{1}{2}))^{2}}].
\end{array}
$$
Since
$$
\begin{array}{rl}
u_{c,t}=ac\left[-(x-ct-\frac{1}{2})
+\frac{\sinh(x-ct-\frac{1}{2})}{2\sinh(\frac{1}{2})}\right]
\end{array}
$$
and
$$
\begin{array}{rl}
u_{c}u_{c,x}
&=a^{2}[\frac{1}{2}(x-ct-\frac{1}{2})^{3}
-(x-ct-\frac{1}{2})^{2}\frac{\sinh(x-ct-\frac{1}{2})}{4\sinh(\frac{1}{2})}
-(x-ct-\frac{1}{2})\frac{\cosh(x-ct-\frac{1}{2})}{2\sinh(\frac{1}{2})}\\[5pt]
&\quad\quad
+\frac{\sinh(x-ct-\frac{1}{2})\cosh(x-ct-\frac{1}{2})}{4(\sinh(\frac{1}{2}))^{2}}
+\frac{47}{24}(x-ct-\frac{1}{2})
-\frac{47}{48}\frac{\sinh(x-ct-\frac{1}{2})}{\sinh(\frac{1}{2})}],
\end{array}
$$
we have
$$
\begin{array}{rl}
\Sigma_{j=1}^{4}I_{j}
=\int_{0}^{T}\int_{\mathbb{S}} a\left[c-(\frac{25}{12}-\frac{\cosh(\frac{1}{2})}{2\sinh(\frac{1}{2})})a\right]\left[-(x-ct-\frac{1}{2})
+\frac{\sinh(x-ct-\frac{1}{2})}{2\sinh(\frac{1}{2})}\right]\varphi dxdt.
\end{array}
$$
Similarly, for $x\leq ct$, we have
$$
\begin{array}{rl}
&u_{c,t}=ac\left[-(x-ct+\frac{1}{2})
+\frac{\sinh(x-ct+\frac{1}{2})}{2\sinh(\frac{1}{2})}\right]\\[5pt]
&u_{c}u_{c,x}=a^{2}[\frac{1}{2}(x-ct+\frac{1}{2})^{3}
-(x-ct+\frac{1}{2})^{2}\frac{\sinh(x-ct+\frac{1}{2})}{4\sinh(\frac{1}{2})}
-(x-ct+\frac{1}{2})\frac{\cosh(x-ct+\frac{1}{2})}{2\sinh(\frac{1}{2})}\\[5pt]
&\qquad\quad\quad~ +\frac{\sinh(x-ct+\frac{1}{2})\cosh(x-ct+\frac{1}{2})}{4(\sinh(\frac{1}{2}))^{2}}
+\frac{47}{24}(x-ct+\frac{1}{2})
-\frac{47}{48}\frac{\sinh(x-ct+\frac{1}{2})}{\sinh(\frac{1}{2})}],
\end{array}
$$
and
$$
\begin{array}{rl}
&g_{x}*[2\mu(u_{c})u_{c}-\frac{1}{2}u_{c,xx}^{2}]
-3g_{xx}*(u_{c,x}u_{c,xx})\\[5pt]
&=a^{2}[-\frac{1}{2}(x-ct+\frac{1}{2})^{3}
+(x-ct+\frac{1}{2})(\frac{1}{8}-\frac{\cosh(\frac{1}{2})}{2\sinh(\frac{1}{2})}
+\frac{\cosh(x-ct+\frac{1}{2})}{2\sinh(\frac{1}{2})})\\[5pt]
&\quad\quad~ +(x-ct)^{2}\frac{\sinh(x-ct+\frac{1}{2})}{4\sinh(\frac{1}{2})}
+(x-ct)\frac{\sinh(x-ct+\frac{1}{2})}{4\sinh(\frac{1}{2})}\\[5pt]
&\quad\quad~ -\frac{\sinh(x-ct+\frac{1}{2})\cosh(x-ct+\frac{1}{2})}{4(\sinh(\frac{1}{2}))^{2}}
+\frac{\sinh(x-ct+\frac{1}{2})\cosh(\frac{1}{2})}{4(\sinh(\frac{1}{2}))^{2}}].
\end{array}
$$
Thus,
$$
\begin{array}{rl}
\Sigma_{j=1}^{4}I_{j}
=\int_{0}^{T}\int_{\mathbb{S}} a\left[c-(\frac{25}{12}-\frac{\cosh(\frac{1}{2})}{2\sinh(\frac{1}{2})})a\right]\left[-(x-ct+\frac{1}{2})
+\frac{\sinh(x-ct+\frac{1}{2})}{2\sinh(\frac{1}{2})}\right]\varphi dxdt.
\end{array}
$$
Since $\varphi$ is arbitrary, both cases imply $a$ satisfies $c-(\frac{25}{12}-\frac{\cosh(\frac{1}{2})}{2\sinh(\frac{1}{2})})a=0$, which completes the proof. \hfill $\Box$

\section*{Acknowledgments}
Wang's work is supported by the Fundamental Research Funds for the Central Universities.
Li's work is supported by NSFC (No:11571057).
Qiao's work is partially supported by the President's Endowed Professorship program of the University of Texas system.

\label{}

%%\appendix


%% References


\bibliographystyle{model3-num-names}
\bibliography{<your-bib-database>}

%% not want to use model3-num-names.bst.


\end{document}